\documentclass[english]{paper}

\usepackage[foot]{amsaddr}
\usepackage{tikz}
\usepackage{tikz-cd}

\usepackage{mama}
\usepackage[utf8]{inputenc}
\DeclareUnicodeCharacter{03C0}{$\pi$}

\usepackage{stmaryrd}
\usepackage{cleveref}
\usepackage{todonotes}
\usepackage{dirtytalk}
\usepackage{stmaryrd}
\usepackage{wrapfig}
\usepackage{listings}
\usepackage{caption}
\usepackage{subcaption}

\usepackage{tikzit}

\tikzstyle{node}=[fill=black, draw=black, shape=circle, scale=0.5]
\tikzstyle{medium_box}=[fill=white, draw=black, shape=rectangle, minimum height=0.8cm, minimum width=0.5cm]

\tikzstyle{pointy}=[->]
\tikzstyle{bluearrow}=[->, fill=none, draw={rgb,255: red,29; green,206; blue,255}, thick]
\tikzstyle{lightnone}=[-, draw={rgb,255: red,191; green,191; blue,191}]

\tikzset{
	Rightarrow/.style={double equal sign distance,>={Implies},->},
	triple/.style={-,preaction={draw,Rightarrow}},
	quadruple/.style={preaction={draw,Rightarrow,shorten >=0pt},shorten >=1pt,-,double,double distance=0.2pt}
}

\newcommand{\GET}{\texttt{GET}}
\newcommand{\POST}{\texttt{POST}}

\usepackage{listings}
\usepackage{lstfiracode} 
\usepackage{newunicodechar}
\newunicodechar{✗}{\times}

\usepackage{fancyvrb}

\makeatletter
\let\FV@ProcessLine\relax
\makeatother

\providecommand{\tightlist}{%
  \setlength{\itemsep}{0pt}\setlength{\parskip}{0pt}}

\DefineVerbatimEnvironment{Highlighting}{Verbatim}{commandchars=\\\{\}}

\title{Lenses for Composable Servers}
\author{Andre Videla, Matteo Capucci}
\email{andre.videla@strath.ac.uk, matteo.capucci@strath.ac.uk}
\address{University of Strathclyde, Glasgow (UK)}

\begin{document}
    \begin{abstract}
        We implement the semantics of server operations using parameterised
        lenses. They allow us to define endpoints and extend them using
        classical lens composition. The parameterised nature of lenses models
        state updates while the lens laws mimic properties expected from HTTP.

        This first approach to server development is extended to use
        dependent parameterised lenses. An upgrade necessary to model
        not only endpoints, but entire servers, unlocking the ability to compose
        them together.
    \end{abstract}

    \maketitle

	\section{Introduction}
Client-server architectures are the bread and butter of modern Internet
applications with HTTP \cite{HTTP} being the underlying technology
enabling them. HTTP allows communication between a
client and a server, and in that context, servers need to be written,
maintained and extended to keep providing services to its
clients. However, the implementation and design of servers have been
relying on ad-hoc conventions \cite{Fielding_2000} and informal practices
to make up for a lack of formal definition of servers.
The goal of this paper is to find a data structure that captures the
essence of web server operations to construct,
extend and compose servers together in an automated manner.

We will see in detail how lenses can be used to fill that role.
Thanks to their use as generic data accessors, we can model a typical server as a
program that gives access to a resource, and lenses focus on different parts of it.
While helpful, that
picture is incomplete because servers need to handle \emph{state} in addition to respond
to requests. To represent state, we are going to rely on a variant of lenses called
``parametrised lenses''\cite{capucci2021towards}, they make use of an additional
``port'' which we use to represent state. From that perspective, servers as
lenses are interaction systems with both an outside client, and a database.
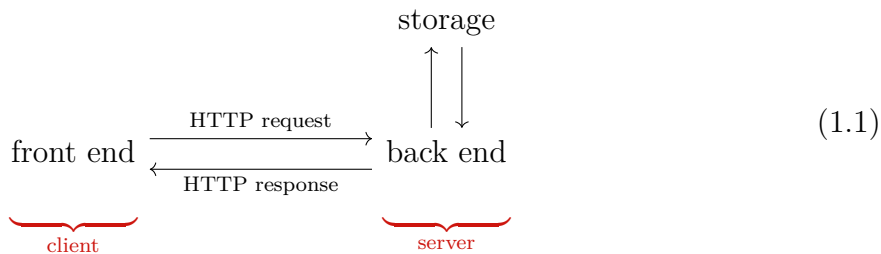
\begin{diagram}
 	&& {\text{storage}} \\
 	{\text{front end}} && {\text{back end}} \\[-6ex]
 	\textcolor{rgb,255:red,205;green,21;blue,14}{\underbrace{\phantom{front end}}_{\text{client}}} && \textcolor{rgb,255:red,205;green,21;blue,14}{\underbrace{\phantom{front end}}_{\text{server}}}
 	\arrow["{\text{HTTP request}}", shift left=2, from=2-1, to=2-3]
 	\arrow[shift left=2, from=2-3, to=1-3]
 	\arrow[shift left=2, from=1-3, to=2-3]
 	\arrow["{\text{HTTP response}}", shift left=2, from=2-3, to=2-1]
 \end{diagram}

We rely heavily on graphical intuition to communicate this correspondence
between server and lenses. And because those ideas have been implemented, we are also
going to rely on code snippets written in Idris \cite{Brady_2013, Brady_2021}. Idris
was picked for its focus on runnable software, in particular its out of the box
support for network primitive while providing the dependent types necessary to
implement dependent lenses.

\hypertarget{contributions}{%
\subsection{Contributions}\label{contributions}}

We provide the following contributions:

\begin{itemize}
\tightlist
\item
  A conceptualisation of resources as lenses.
\item
  A definition of servers as parametrised dependent lenses.
\item
  An implementation of a server library using the preceding
  contributions.
\end{itemize}

Those contributions will guide the structure of this document. In
the first part, Section~\ref{polymorphic-parametrised-lenses}, we introduce
plain and parametrised lenses. A reader familiar with them can skip it
and jump directly to Section~\ref{resources-as-parametrised-lenses}
where we introduce \emph{resources} as parametrised lenses. Resources
for web servers are traditionally documents served to a client. But
with more dynamical web pages, resources come from a database, and
clients ask the server to provide or modify the data living on the
database. We are going to view resources as remote data that we can
both query and update. In that interpretation, \emph{servers act as lenses
focusing on different parts of their internal storage}. This suggests
that we can construct them using the same bulding blocks used for lenses,
unlocking the ability to use the existing ecosystem of lens libraries
to build servers.

We notice that lenses are not enough to build servers because they are
stateless, whereas servers perform state manipulation, for this we
move to parametrised lenses, whose parameter will represent our state.
This implies that the server mediates the communication between the
client and its state.

While useful, this construction hits a roadblock when we attempt
to combine multiple lenses together. In order to build a server we
need to step outside of the realm of lenses and write
a server engine that perform routing and handling.

To solve this issue we introduce \emph{dependent lenses} and
\emph{dependent parametrised lenses} in Section~\ref{dependent-parametrised-lenses}. They provide us with two new
operators that perform the coproduct on parametrised lenses.

We will use this coproduct operation in
Section~\ref{server-as-dependent-lenses} to combine endpoints into
larger servers and even combine servers together. Finally,
in the last section, Section~\ref{the-recombine-library}, we look at the
benefits of implementing these ideas in a programming
language by presenting \textsc{recombine} a server library written in
Idris built upon parametrised dependent lenses and providig an embedded
domain specific language to write servers.

\hypertarget{polymorphic-parametrised-lenses}{%
\section{Polymorphic (parametrised) lenses}\label{polymorphic-parametrised-lenses}}

Lenses in functional programming have the following structure:

\begin{itemize}
\tightlist
\item
  A \texttt{view} function that retrieves the information nested inside
  a structure.
\item
  An \texttt{update} function that returns a new version of the overall
  structure after modification using an argument.
\end{itemize}

Their original intent was to explain view and update in programs such as
databases \cite{Foster_Greenwald_Moore_Pierce_Schmitt_2007}.
In practice, they gained popularity in the context of
\emph{pure functional programming} where
mutating state is impossible, yet operations to update nested data are still
necessary to implement real-world functionality. Lenses have
been implemented into multiple libraries (such as \cite{kmett-lens,quicklens})
to allow seamless manipulation of nested data by providing operators to compose lenses
and use them to query objects. More recently, lenses have
been extensivelty generalized, to so-called \emph{optics}, and studied formally using advanced category theory \cite{bryce2020categoricalupdate}.
Additionally, optics (and therefore lenses) have been found very helpful in modeling interactive
systems such as games and machine learning models \cite{towards_foundations_categorical_cybernetics, bolt_hedges_zahn_bayesian_open_games}
which suggests they are not only suited for data access, but also for dynamical
systems.

This section will recap the definition of lenses we use and
highlight the parts that are relevant for our servers. The code samples will be
written in Idris, because the library we present in Section~\ref{the-recombine-library}
is written in Idris. Idris' syntax should be familiar to readers accustomed
with Haskell, though we are going to depart from some of the traditional notation.
In particular, products (pairs) are written using \texttt{(*)} at the type level to
distinguish them from their constructors at the term-level: \texttt{(,)}. Coproducts
(the \texttt{Either} type) will be written using \texttt{(+)} at the type level to be
symetric with \texttt{(*)}. Idris features \emph{records} as well as \emph{dot postfix
notation} which mimic syntax from object-oriented programming languages. We will make
use of both of them to streamline the use of field projections.

A very small example of lenses is a record for a \texttt{User} which contains
a field for an \texttt{Address}, which itself contains fields for its street
name, street number and city.

\begin{Verbatim}[fontsize=\small]

record Address where
  constructor MkAddress
  city : String
  streetName : String
  streetNumber : Int

record User where
  constructor MkUser
  username : String
  address : Address
  birthdate : Date
\end{Verbatim}

Trying to update the street number of a user by copying the fields results in
unsatisfying code gymnastics:

\begin{Verbatim}[fontsize=\small]
changeStreetName : String -> User -> User
changeStreetName newName (MkUser un (MkAddress c oldName num) bd) =
  MkUser un (MkAddress newName c num) bd
\end{Verbatim}

The entire constructor needs to be destructured and copied, save for the
new value. But as this example shows, not only it's tedious, it's also
error prone since neither the syntax nor the types allow us to catch the
fact that we've flipped the street name and the city in the example.

Lenses aim to solve this problem of composing nested updates by abstracting
the projection and update functions into their own records.
A classical implementation for lenses would be:

\begin{Verbatim}[fontsize=\small]
record Lens (s, t, a, b : Type) where
  constructor MkLens
  view : s -> a
  update : s -> b -> t
\end{Verbatim}

The \texttt{view} returns a sub-part of the data. Conversely, given an argument \texttt{b},
which represents a \emph{change} in the data, the \texttt{update} returns new data.
The true power of the lens abstraction, however, is in encapsulating these operation into composable and reusable blocks.
Indeed, assuming lenses for each field of our previous example, our change in street name is
expressed with the following program:

\begin{Verbatim}[fontsize=\small]
changeStreetName : User -> Int -> User
changeStreetName = update (userLens |> streetLens)
\end{Verbatim}

where \texttt{(|>)} performs lens composition, which we will define in Section~\ref{composition}.

To better capture the expected behavior of lenses, sometimes one limits themselves to so-called \emph{monomorphic lawful lenses}, which are lenses such that $s=t$ and $a=b$ (this is what \emph{monomorphic} means) and that moreover follow a set of laws:

\begin{Verbatim}[fontsize=\small]
-- You get back what you've put in
put-get : forall x, v, lens. lens.view (lens.update x v) = v
-- Putting things in twice in a row is the same as doing it once
put-put : forall x, v, lens. lens.update (lens.update x v) v
        = lens.update x v
-- if you put in what you see you won't change anything
get-put : forall v, lens. lens.update (lens.view v) v = v
\end{Verbatim}

It's very important that these properties are stable under composition, that is compositions of lawful lenses are still lawful \cite{Riley_2018}.

Throughout this paper we will use three representations for lenses:

\hypertarget{lenses-as-morphisms-between-boundaries}{%
\subsubsection{Lenses as morphisms between boundaries}\label{lenses-as-morphisms-between-boundaries}}

This notation exposes the nature of lenses as morphisms between boundaries and is written as $(X, S) \to (Y, R)$.
They decompose into functions $X \to Y$ which we call the \emph{forward} part, and $X \to R \to S$, which
we call the \emph{backward} part.

\hypertarget{lenses-as-string-diagrams}{%
\subsubsection{Lenses as string diagrams}\label{lenses-as-string-diagrams}}

\begin{figure}
\begin{center}
\begin{tikzpicture}

\draw (0.5,0) rectangle (3.5,2);
    \draw[->] (0, 1.5) node [left] {X} -- (0.5,1.5) ;
    \draw[<-] (0, 0.5) node [left] {S} -- (0.5,0.5) ;
    \draw[->] (3.5, 1.5) -- (4.0,1.5) node [right] {Y} ;
    \draw[<-] (3.5, 0.5) -- (4.0,0.5) node [right] {R} ;

    \draw (1.5,1.3 ) rectangle (2.5,1.7) ;
    \draw (2, 1.5) node {a1} ;
    \draw [->] (0.5,1.5) -- (1.5,1.5) ;
    \draw (2.5,1.5) -- (3.5,1.5) ;

    \draw (1.5,0.3) rectangle (2.5,0.7) ;
    \draw (2, 0.5) node  {a2} ;
    \draw (0.5,0.5) -- (1.5,0.5) ;
    \draw [<-] (2.5,0.5) -- (3.5,0.5) ;

\end{tikzpicture}
\end{center}
\caption{An adapter between boundaries $(X, S)$, $(Y, R)$}
\label{adapter-lens-diagram}
\end{figure}
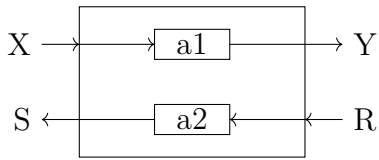

We can represent lenses as black boxes with left and right boundaries that
compose together. The composition operations will follow shortly.
Sometimes it is informative to draw the content of the lens to
understand how they are implemented and how the flow of data is directed
through them. You see in Figure~\ref{poly-lens-diagram} a typical lens with
a view and update function, the top left input $(X)$ is shared between the
\texttt{view} and \texttt{update}. The direction of arrows shows the flow
of data through the lens and from and explains why the \texttt{view} function
is called \emph{forward} and \texttt{update} is called \emph{backward}
(arrows flow from right to left).
In Figure~\ref{adapter-lens-diagram} the
lens is an \emph{adapter}, adapters do not share the top left
input with the backward part.

\begin{figure}
\begin{center}
\begin{tikzpicture}

\draw (0.5,0) rectangle (3.5,2);
    \draw[->] (0, 1.5) node [left] {X} -- (0.5,1.5) ;
    \draw[<-] (0, 0.5) node [left] {S} -- (0.5,0.5) ;
    \draw[->] (3.5, 1.5) -- (4.0,1.5) node [right] {Y} ;
    \draw[<-] (3.5, 0.5) -- (4.0,0.5) node [right] {R} ;

    \draw (2.2,1.3 ) rectangle (3.2,1.7) node [anchor=north east, yshift=2] {\small view} ;
    \draw [->] (0.5,1.5) -- (2.2,1.5) ;
    \draw (3.2,1.5) -- (3.5,1.5) ;

    \draw [->](2,1.5) node[circle,fill,inner sep=1pt] {} -- (2,0.7)  ;

    \draw (1.3,0.3) rectangle (2.7,0.7) ;
    \draw (2.0,0.5) node {\small update} ;
    \draw (0.5,0.5) -- (1.3,0.5) ;
    \draw [<-] (2.7,0.5) -- (3.5,0.5) ;

\end{tikzpicture}
\end{center}
\caption{A lens with boundaries $(X, S)$, $(Y, R)$}
\label{poly-lens-diagram}
\end{figure}
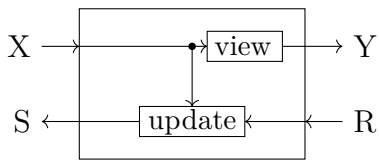

\hypertarget{lenses-as-programs}{%
\subsubsection{Lenses as programs}\label{lenses-as-programs}}

Lenses can be implemented as a product of two functions, one for
the forward and one for the backward part of the lens.
While we could use the implementation given earlier, we choose to represent
boundaries explicitly rather than rely on four unrelated type
parametrs.
We achieve this in Idris by declaring lenses as a record
parametrised over two \emph{boundaries}:

\begin{Verbatim}[fontsize=\small]
Boundary : Type
Boundary = Type * Type

record Lens (l, r : Boundary) where
    constructor MkLens
    view : l.π1 -> r.π1
    update : l.π1 -> r.π2 -> l.π2
\end{Verbatim}

Using boundaries instead of type parameters recalls their notation as
morphism between boundaries, though its functions need to be
defined in terms of projections of the boundaries where
\texttt{π1\ :\ a * b\ -\textgreater{}\ a} and
\texttt{π2\ :\ a * b\ -\textgreater{}\ b}.

\hypertarget{operations-on-lenses}{%
\subsection{Operations on lenses}\label{operations-on-lenses}}

As anticipated, the strength of lenses lies in their compositional properties.
The two operations that we will use on lenses are \emph{sequential composition}
and \emph{parallel composition}. The first one stitches two lenses together one
after the other, assuming their connecting boundaries are the same. For data
accessors, this is used to focus on deeper parts of a record.
Instead, parallel composition is used to focus on
two different parts of a record simultaneously.

\hypertarget{composition}{%
\subsubsection{Composition}\label{composition}}

Sequential composition chains two lenses together when their boundaries
agree. The first lens must have a right boundary that is compatible with
the left boundary of the second lens. This means composition is a map like this
\[
    \vartriangleright : \left((X, T) \to (Y, S)\right) \times \left((Y, S) \to (Z, R)\right) \longto (X, T) \to (Z, R).
\]

\begin{Verbatim}[fontsize=\small]
(|>) : (a : Lens l x) -> (b : Lens x r) -> Lens l r
(|>) (MkLens v1 u1) (MkLens v2 u2) = MkLens
  (v2 . v1)
  (\st, val => u1 st (u2 (v1 st) val))
\end{Verbatim}


\begin{figure}
\begin{tikzpicture}

\draw (0.5,0) rectangle (3.5,2);
\draw[->] (0, 1.5) node [left] {X} -- (0.5,1.5) ;
\draw[<-] (0, 0.5) node [left] {T} -- (0.5,0.5) ;
\draw[->] (3.5, 1.5) -- (4.0,1.5) ;
\draw[<-] (3.5, 0.5) -- (4.0,0.5) ;
\draw (5.0,0) rectangle (8.0,2);
\draw[->] (8, 1.5) -- (8.5,1.5) node [right] {Z};
\draw[<-] (8, 0.5) -- (8.5,0.5) node [right] {R};
\draw[->] (4.5, 1.5) node [left] {Y}  -- (5.0,1.5) ;
\draw[<-] (4.5, 0.5) node [left] {S}  -- (5.0,0.5) ;

\end{tikzpicture}
    \caption{Composition of lenses with boundaries $(X, T) \to (Y, S)$ and $(Y, S) \to (Z, R)$}
    \label{compose-lens-diagram}
\end{figure}
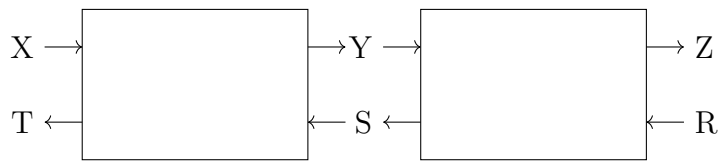

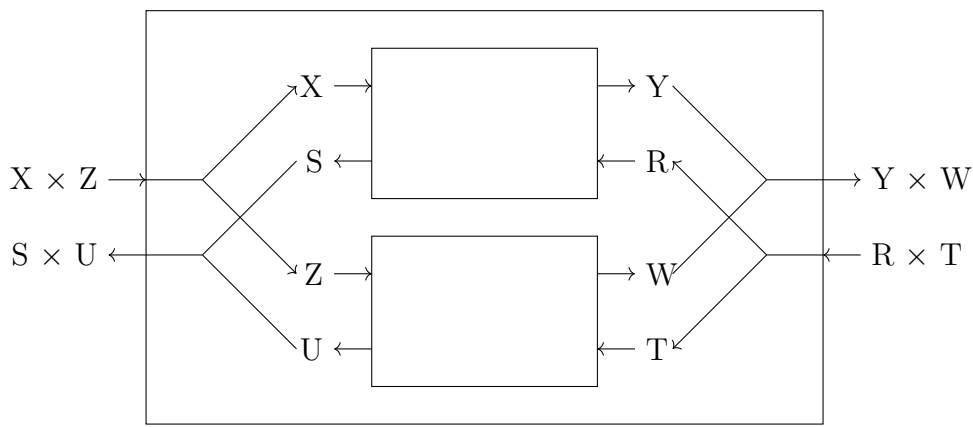
\begin{figure}
\begin{tikzpicture}

\draw (2,2.5) rectangle (5,4.5);
\draw[->] (1.5, 4) node [left] {X} -- (2,4) ;
\draw[<-] (1.5, 3) node [left] {S} -- (2,3) ;
\draw[->] (5, 4) -- (5.5,4) node [right] {Y} ;
\draw[<-] (5, 3) -- (5.5,3) node [right] {R} ;

    \draw [<-] (1.0,4) -- (-0.25, 2.75) ;
    \draw [->] (-0.25, 2.75) -- (1.0, 1.5) ;

    \draw (2,0) rectangle (5,2);
\draw[->] (1.5, 1.5) node [left] {Z}  -- (2,1.5) ;
\draw[<-] (1.5, 0.5) node [left] {U}  -- (2,0.5) ;
\draw[->] (5, 1.5) -- (5.5,1.5) node [right] {W};
\draw[<-] (5, 0.5) -- (5.5,0.5) node [right] {T};

    \draw (1.0,3) -- (-0.25, 1.75) ;
    \draw (-0.25, 1.75) -- (1.0, 0.5) ;

\draw[<-] (-1, 2.75) -- (-1.5,2.75) node [left] {X $\times$ Z} ;
\draw[<-] (-1.5, 1.75) node [left] {S $\times$ U}  -- (-1,1.75) ;
\draw (-1, 2.75) -- (-0.25,2.75) ;
\draw (-1, 1.75) -- (-0.25,1.75) ;

    \draw [<-] (6.0,3) -- (7.25, 1.75) ;
    \draw [<-] (6.0, 0.5) -- (7.25, 1.75) ;
    \draw (6.0,4) -- (7.25, 2.75) ;
    \draw (6.0, 1.5) -- (7.25, 2.75) ;
    \draw (7.25, 1.75) -- (8, 1.75) ;
    \draw (7.25, 2.75) -- (8, 2.75) ;
    \draw [<-] (8, 1.75) -- (8.5, 1.75) node [right] {R $\times$ T};
    \draw [->] (8, 2.75) -- (8.5, 2.75) node [right] {Y $\times$ W};

\draw (-1, -0.5) rectangle (8, 5);

\end{tikzpicture}
    \caption{Parallel composition of two lenses with boundaries $(X, S) \to (Y, R)$ and $(Z, U) \to (W, T)$}
    \label{parallel-composition-diagram}
\end{figure}

\hypertarget{parallel-composition}{%
\subsubsection{Parallel composition}\label{parallelcomposition}}

The \emph{parallel} composition of two lenses produces a lens
that focuses on two distincts parts of a product.
In order to properly combine the boundaries, we perform the
product on boundaries that we define as the point-wise product of its
components.

\begin{Verbatim}[fontsize=\small]
-- multiplication of boundaries
(*) : Boundary -> Boundary -> Boundary
(x, y) * (a, b) = (x * a, y * b)

parallel : Lens l1 r1 -> Lens l2 r2 -> Lens (l1 * l2) (r1 * r2)
parallel (MkLens v1 u1) (MkLens v2 u2) = MkLens
  (bimap v1 v2)
  (\(s1, s2), (b1, b2) => (u1 s1 b1, u2 s2 b2))
\end{Verbatim}

\hypertarget{parametrised-lenses}{%
\subsection{Parameterised lenses}\label{parametrised-lenses}}

The final piece of technology we need to introduce is \emph{parametrised lenses}. They are lenses
augmented with an additional boundary on the top. This version of lenses arises by applying the $\mathrm{Para}$ construction on lenses, as detailed in \cite{capucci2021towards}. For our purposes, the parametr
exposes the state of the server. Adding a \emph{parameter} to our lenses changes the
nature of the lens algebra allowing the composition of
lenses along an extra dimension.
Effectively, the parameter boundary is attached to the left boundary using the product of
boundaries outlined above:
\[
    (P, Q) \times (X, S) \to (Y, R)
\]

\begin{Verbatim}[fontsize=\small]
ParaLens : Boundary -> Boundary -> Boundary -> Type
ParaLens left para right = Lens (left * para) right
\end{Verbatim}

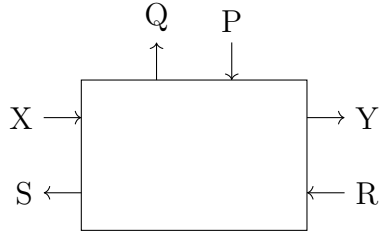
\begin{figure}[H]
    \centering
\begin{tikzpicture}

\draw (0.5,0) rectangle (3.5,2);
\draw[->] (0, 1.5) node [left] {X} -- (0.5,1.5) ;
\draw[<-] (0, 0.5) node [left] {S} -- (0.5,0.5) ;
    \draw[->] (3.5, 1.5) -- (4.0,1.5) node [right] {Y} ;
    \draw[<-] (3.5, 0.5) -- (4.0,0.5) node [right] {R} ;
    \draw [->] (1.5, 2) -- (1.5, 2.5) node [above] {Q} ;
    \draw [<-] (2.5, 2) -- (2.5, 2.5) node [above] {P} ;
\end{tikzpicture}
\end{figure}

From now on, we refer non-parametrised lenses as
\emph{plain lenses}. In what follows we explain operators on parametrised
lenses.

\hypertarget{embedding-plain-lenses}{%
\subsubsection{Embedding plain lenses}\label{embedding-plain-lenses}}

We observe that plain lenses can be interpreted
as parametrised lenses with a \texttt{unit} top boundary:
$\text{toPara }((X, S) \to (Y, R)) = ((), ()) \times (X, S) \to (Y, R)$

\begin{Verbatim}[fontsize=\small]
toPara : Lens l r -> Para l ((), ()) r
toPara lens =
  MkLens (lens.view . π1) (\x => (,()) . lens.update x.π1)
\end{Verbatim}

\hypertarget{composition-of-parametrised-lenses}{%
\subsubsection{Composition of parametrised lenses}\label{composition-of-parametrised-lenses}}

The addition of a parameter seems innocuous but it changes composition. Composed
parametrised lenses combine their parameter with a product, that means that the
resulting lens has access to the state of both lenses at once.


\begin{Verbatim}[fontsize=\small]
(|>) : Para l p1 x -> Para x p2 r ->
       Para l (p1 * p2) r
\end{Verbatim}

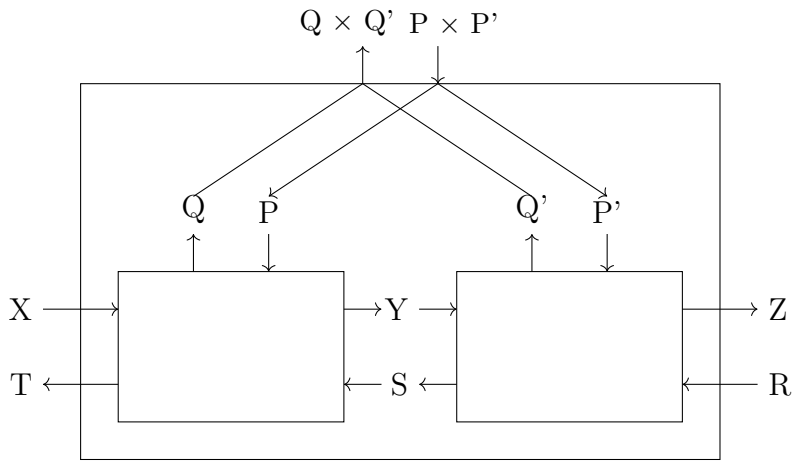
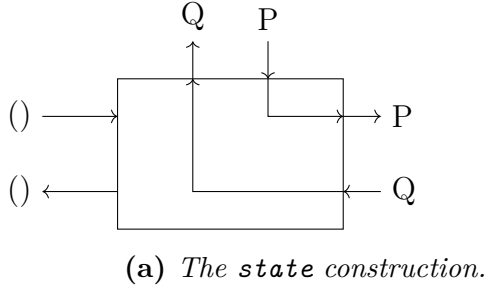
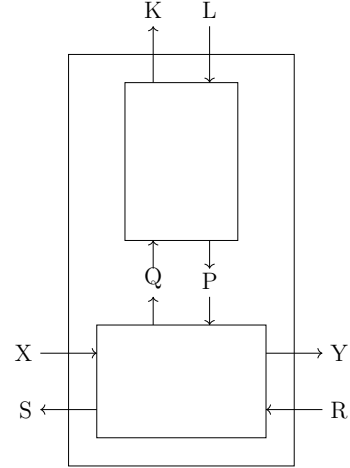
\begin{figure}[H]
    \centering
\begin{tikzpicture}

\draw (0.5,0) rectangle (3.5,2);
\draw[->] (-0.5, 1.5) node [left] {X} -- (0.5,1.5) ;
\draw[<-] (-0.5, 0.5) node [left] {T} -- (0.5,0.5) ;
\draw[->] (3.5, 1.5) -- (4.0,1.5) ;
\draw[<-] (3.5, 0.5) -- (4.0,0.5) ;
\draw (5.0,0) rectangle (8.0,2);
\draw[->] (8, 1.5) -- (9,1.5) node [right] {Z};
\draw[<-] (8, 0.5) -- (9,0.5) node [right] {R};
\draw[->] (4.5, 1.5) node [left] {Y}  -- (5.0,1.5) ;
\draw[<-] (4.5, 0.5) node [left] {S}  -- (5.0,0.5) ;

    \draw [->] (6, 2) -- (6, 2.5) node [above] {Q'} ;
    \draw [<-] (7, 2) -- (7, 2.5) node [above] {P'};
    \draw [->] (1.5, 2) -- (1.5, 2.5) node [above] {Q} ;
    \draw [<-] (2.5, 2) -- (2.5, 2.5) node [above] {P} ;

    \draw [->] (3.75, 4.5) -- (3.75, 5.0) node [above, xshift=-6] {\small Q $\times$ Q'};
    \draw [<-] (4.75, 4.5) -- (4.75, 5.0) node [above, xshift=6] {\small P $\times$ P'};
    \draw (1.5, 3) -- (3.75, 4.5) ;
    \draw (6, 3) -- (3.75, 4.5) ;

    \draw [<-] (2.5, 3) -- (4.75, 4.5) ;
    \draw [<-] (7, 3) -- (4.75, 4.5) ;

    \draw (0, -0.5) rectangle (8.5,4.5);
\end{tikzpicture}
\end{figure}

\hypertarget{the-state-lens}{%
\subsubsection{The state lens}\label{the-state-lens}}

Using `corner wires' (see Figure~\ref{fig:state}) we can expose the top boundary as the focus of our lens for querying and updating.
In our server interpretation, the top boundary will play the role of state
parametrs, that is why we call this lens \texttt{State}.

\begin{Verbatim}[samepage=true]
state : (b : Boundary) -> Para ((), ()) b b
state (MkB p q) = MkLens snd (const ((),))
\end{Verbatim}

\hypertarget{reparametrisation}{%
\subsubsection{Reparametrisation}\label{reparametrisation}}

Reparametrisation (Figure~\ref{fig:reparam}) allows changing the boundary along the top using a plain lens,
this is the extra dimension along which we can compose lenses.


\begin{Verbatim}[fontsize=\small]
reparam : Para l p r -> Lens p' p -> Para l p' r
\end{Verbatim}

\begin{figure}
    \begin{subfigure}{0.5\textwidth}
\begin{tikzpicture}
\draw (0.5,0) rectangle (3.5,2);
    \draw[->] (-0.5, 1.5) node [left] {()} -- (0.5,1.5) ;
    \draw[<-] (-0.5, 0.5) node [left] {()} -- (0.5,0.5) ;
    \draw[->] (3.5, 1.5) -- (4.0,1.5) node [right] {P} ;
    \draw[<-] (3.5, 0.5) -- (4.0,0.5) node [right] {Q} ;
    \draw [->] (1.5, 2) -- (1.5, 2.5) node [above] {Q} ;
    \draw [<-] (2.5, 2) -- (2.5, 2.5) node [above] {P} ;


    \draw[->] (2.5,2) -- (2.5,1.5) -- (3.5,1.5);
    \draw[<-] (1.5,2) -- (1.5,0.5) -- (3.5,0.5);
\end{tikzpicture}
\caption{The \texttt{state} construction.}
\label{fig:state}
\hfill
\end{subfigure}
    \begin{subfigure}{0.45\textwidth}
    \scalebox{0.75}{
\begin{tikzpicture}
\draw (1,3.5) rectangle (3,6.3);
    \draw [->] (1.5, 6.3) -- (1.5, 7.3) node [above] {K} ;
    \draw [<-] (2.5, 6.3) -- (2.5, 7.3) node [above] {L} ;
    \draw [->] (1.5, 3) -- (1.5, 3.5) ;
    \draw [<-] (2.5, 3) -- (2.5, 3.5) ;

\draw (0.5,0) rectangle (3.5,2);
\draw[->] (-0.5, 1.5) node [left] {X} -- (0.5,1.5) ;
\draw[<-] (-0.5, 0.5) node [left] {S} -- (0.5,0.5) ;
    \draw[->] (3.5, 1.5) -- (4.5,1.5) node [right] {Y} ;
    \draw[<-] (3.5, 0.5) -- (4.5,0.5) node [right] {R} ;
    \draw [->] (1.5, 2) -- (1.5, 2.5) node [above] {Q} ;
    \draw [<-] (2.5, 2) -- (2.5, 2.5) node [above] {P} ;
    \draw (0,-0.5) rectangle (4.0, 6.8) ;
\end{tikzpicture}
    }
    \caption{Reparametrisation of a lens changing the top boundary from $(P, Q)$ to $(K, L)$.}
    \label{fig:reparam}
    \end{subfigure}
\end{figure}

\hypertarget{pre-compositionb}{%
\subsubsection{Pre-composition $(\lll)$}\label{pre-composition}}

Pre-composition can be implemented in terms of
\ref{embedding-plain-lenses}, \ref{composition-of-parametrised-lenses}, and \ref{reparametrisation}.
It allows changing the left boundary using a plain lens. This will be
used to pre-process incoming requests before handling them.

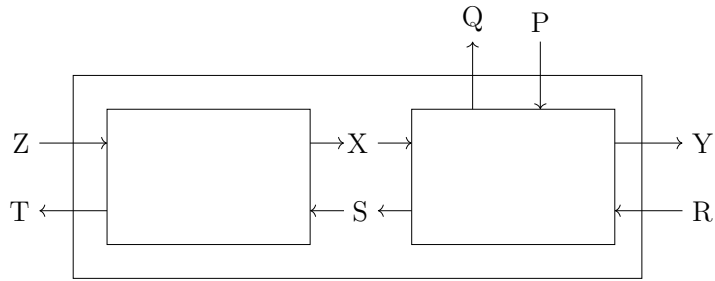
\begin{Verbatim}[fontsize=\small]
(<<<) : Lens l x -> Para x p r -> Para l p r
\end{Verbatim}

\begin{figure}[H]
\scalebox{0.9}{%
\begin{tikzpicture}

\draw (0.5,0) rectangle (3.5,2);
\draw[->] (-0.5, 1.5) node [left] {Z} -- (0.5,1.5) ;
\draw[<-] (-0.5, 0.5) node [left] {T} -- (0.5,0.5) ;
\draw[->] (3.5, 1.5) -- (4,1.5) ;
\draw[<-] (3.5, 0.5) -- (4,0.5) ;

    \draw[->] (5.9, 2) -- (5.9, 3) node (0)[above] {Q};
    \draw[<-] (6.9, 2) -- (6.9, 3) node (1)[above] {P};

\draw (5.0,0) rectangle (8.0,2);
\draw[->] (4.5, 1.5) node [left] {X}  -- (5.0,1.5) ;
\draw[<-] (4.5, 0.5) node [left] {S}  -- (5.0,0.5) ;
\draw[->] (8.0, 1.5) -- (9,1.5) node [right] {Y};
\draw[<-] (8.0, 0.5) -- (9,0.5) node [right] {R};

    \draw (0,-0.5) rectangle (8.4, 2.5);

\end{tikzpicture}
}
\end{figure}
\hypertarget{post-compositionb}{%
\subsubsection{Post-composition $(\ggg)$}\label{post-composition}}
Similarily with post-composition, it allows changing the right boundary
using a plain lens:
This will be used to implement new endpoints from existing ones by
focusing on deeper parts of a resource.
\begin{Verbatim}[fontsize=\small]
(>>>) : Para l p x -> Lens x r ->  Para l p r
\end{Verbatim}

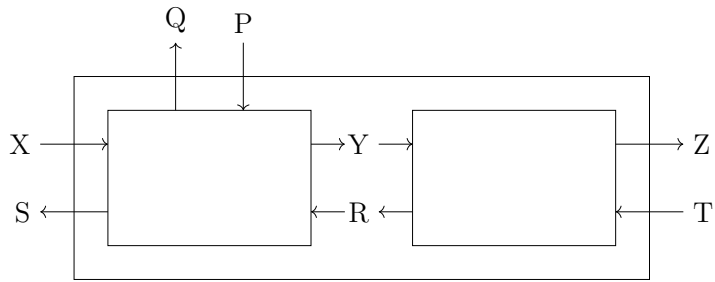
\begin{figure}[H]
    \centering
\scalebox{0.9}{%
\begin{tikzpicture}

\draw (0.5,0) rectangle (3.5,2);
\draw[->] (-0.5, 1.5) node [left] {X} -- (0.5,1.5) ;
\draw[<-] (-0.5, 0.5) node [left] {S} -- (0.5,0.5) ;
\draw[->] (3.5, 1.5) -- (4,1.5) ;
\draw[<-] (3.5, 0.5) -- (4,0.5) ;

    \draw[->] (1.5, 2) -- (1.5, 3) node (0)[above] {Q};
    \draw[<-] (2.5, 2) -- (2.5, 3) node (1)[above] {P};

\draw (5.0,0) rectangle (8.0,2);
\draw[->] (4.5, 1.5) node [left] {Y}  -- (5.0,1.5) ;
\draw[<-] (4.5, 0.5) node [left] {R}  -- (5.0,0.5) ;
\draw[->] (8.0, 1.5) -- (9,1.5) node [right] {Z};
\draw[<-] (8.0, 0.5) -- (9,0.5) node [right] {T};

    \draw (0,-0.5) rectangle (8.5, 2.5);

\end{tikzpicture}

}
\end{figure}

    \hypertarget{resources-as-parametrised-lenses}{%
\section{Resources as parametrised lenses}\label{resources-as-parametrised-lenses}}

Our first contribution consists in relating \emph{lenses} and resources.
We defined a resource on a server as being a pair of endpoints: one \GET
endpoint to retrieve data and a \POST endpoint to update the same data.
This section will explain how we go from pairs of such endpoints to
server implementation using lenses.
 We start by recapping the properties of HTTP we are
interested in modeling, then explain how those properties can be
captured as \emph{plain lenses} and finally we move on to implementing
servers using \emph{parametrised lenses}.

\hypertarget{http}{%
\subsection{HTTP}\label{http}}
Servers communicate
through HTTP\cite{HTTP}, where requests are
sent from the client to the server and the server responds with the
content asked. The use of lenses and resources is inspired by \emph{RESTful}
servers architecture \cite{Fielding_2000} which is widely used in
commercial software.
HTTP requests are composed of many parts: HTTP method, URI, protocol
version, headers, and body. For our purposes, we are going to direct our
attention to HTTP methods, URI, and request body. Here are the three things to
look out for:

\hypertarget{http-methods}{%
\subsubsection{HTTP Methods}\label{http-methods}}

HTTP methods have different properties, we have laid them out in the
following table:\\

\begin{center}
\begin{tabular}{l|c|c|l}
    Method & Side effects & Idempotent & Description\\
\hline
    GET & NO & YES & Obtain a value from the server.\\
    POST & YES & NO & Change some data on the server.\\
    DELETE & YES & YES & Remove some data from the server.\\
    PUT & YES & YES & Replace some data on the server.\\
    PATCH & YES & NO & Append some data on the server.
\end{tabular}\\
\end{center}

The table shows that endpoints that use HTTP methods such as \texttt{PUT},
\texttt{GET} and \texttt{DELETE} are meant to be idempotent, but others
do not have to.
\hypertarget{uri}{%
\subsubsection{URI}\label{uri}}

Each resource can be accessed through its \emph{URI} which
traditionally represented the path to the file on the server's
filesystem. Nowadays, due to the dynamic nature of web pages, servers
parse the URI and use it to decide which resource to return irrespective
of its path in the filesystem and usually from a database.
Additionally, URIs can carry data using
\emph{captures} e.g: \texttt{/user/:id/name}

This means that paths such as \texttt{/user/3/name} are routed through this api. But
\texttt{/user/book} will not, because ``book'' is not an ID.
In this paper we are going to write down the \emph{type} that we expect our captures
to have, this way, if we expect an \texttt{Int} as id, we write \texttt{/user/Int:id/name}.
URI and captures being such a fundamental feature of HTTP requests, our
server interpretation must have a way to represent them.

\hypertarget{request-body}{%
\subsubsection{Request body}\label{request-body}}

The request body is an arbitrary blob of data that is parsed by the
server. In recent years, this data is often serialised to JSON.
\texttt{GET} requests are special because their
request body is not meant to carry any information.

\hypertarget{servers-as-functions}{%
\subsection{Servers as functions}\label{servers-as-functions}}

After getting familiar with the conditions under which a server must
operate we can start building abstractions to represent them.
A functional programmer's way
to do this would be to lay out the functions which operate a server and
identify their types and their properties, so we are going to do just
that for the endpoints of a server.

We are going to focus on \GET and \POST endpoints since \POST can
implement every other HTTP method by restricting it. And \GET
requests are the only ones that do not perform a side effect on
the server and do not carry a request body.
If we recall the lens laws from Section~\ref{polymorphic-parametrised-lenses} we see that the \texttt{put-put} law ensures
that performing an update twice in a row produces the same result as
updating once, encoding idempotency of update.
This suggests that lawful lenses encode \texttt{PUT} or \texttt{DELETE}
requests while unlawful lenses encode \POST or \texttt{PATCH} requests.

\hypertarget{get-endpoints-as-the-client}{%
\subsubsection{GET endpoints for a client}\label{get-endpoints-as-the-client}}

From the point of view of a client, \texttt{GET} endpoints take as
argument the content of an HTTP request and return some resource from
the server.

\begin{Verbatim}[fontsize=\small]
getEndpoint : Request -> Response
\end{Verbatim}

More precisely, the arguments that the GET endpoint has access to are
the URI, since the body is supposed to carry no information whatsoever.

\begin{Verbatim}[fontsize=\small]
getEndpoint : uri -> Response
\end{Verbatim}

Finally, the \texttt{Reponse} type is a bit too broad, each endpoint
returns a single value which will be serialised in the body of the
response. The rest of the response will be built the same way for each
endpoint. After refactoring the common parts we are left with:

\begin{Verbatim}[fontsize=\small]
getEndpoint : Serialisable resultGet => uri -> resultGet
\end{Verbatim}

\hypertarget{post-endpoints-as-the-client}{%
\subsubsection{POST endpoints for a
client}\label{post-endpoints-as-the-client}}

Unlike \texttt{GET} endpoints, \texttt{POST} endpoints make essential use of their request
body and also modify the state of the server, so while the overall
picture \texttt{postEndpoint:\ Request\ -\textgreater{}\ Response} is
the same, refining the types results in the following:

\begin{Verbatim}[fontsize=\small]
postEndpoint : Parsable body => Serialisable resultPost =>
               uri -> body -> resultPost
\end{Verbatim}

We have not addressed the fact that this function can change the state
of the server, but from the point of view of the client, this is
irrelevant, the client only cares about the result given and will trust
that the server performed the side effect as expected.

\hypertarget{resources-as-lenses}{%
\subsection{Resources as lenses}\label{resources-as-lenses}}

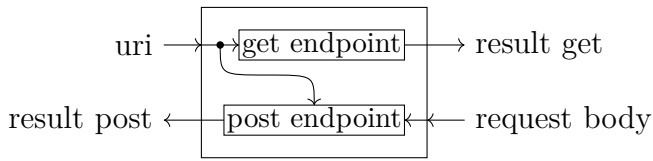
\begin{figure}
\begin{tikzpicture}

\draw (0.5,0) rectangle (3.5,2);
\draw[->] (0, 1.5) node [left] {uri} -- (0.5,1.5) ;
\draw[<-] (0, 0.5) node [left] {result post} -- (0.5,0.5) ;
    \draw[->] (3.5, 1.5) -- (4.0,1.5) node [right] {result get};
    \draw[<-] (3.5, 0.5) -- (4.0,0.5) node [right] {request body};

    \draw (1,1.3) rectangle (3.2,1.7) ;
    \draw (2.1,1.5) node {\small get endpoint} ;
    \draw [->] (0.5,1.5) -- (1,1.5) ;
    \draw (3.2,1.5) -- (3.5,1.5) ;
    \draw [->](0.75,1.5) node[circle,fill,inner sep=1pt] {} .. controls (0.75, 1.1) .. (1.375, 1.1) .. controls (2, 1.1) .. (2,0.7) ;
    \draw (0.8,0.3) rectangle (3.2,0.7) ;
    \draw (2, 0.5) node {\small post endpoint} ;
    \draw (0.5,0.5) -- (0.8,0.5);
    \draw [<-] (3.2,0.5) -- (3.5,0.5);

\end{tikzpicture}
    \caption{A resource as a plain lens}
    \label{resource-as-plain-lens}
\end{figure}

Earlier, we defined pairs of endpoints viewing and updating the same data
as a \emph{resource}.
Placing the two functions describing our resource side by side yields
something very close to a lens:

\begin{Verbatim}[fontsize=\small]
getEndpoint : Serialisable resultGet => Parsable uri =>
              uri -> resultGet
postEndpoint : Parsable body => Serialisable resultPost =>
               uri -> body -> resultPost
\end{Verbatim}

In fact, if we remove the constraints we see in Figure~\ref{resource-as-plain-lens}
that \texttt{uri\ -\textgreater{}\ resultGet} forms the forward part of a
lens, while
\texttt{uri\ -\textgreater{}\ body\ -\textgreater{}\ resultPost} is the
backward part.

If we assume our server is composed of two endpoints that permit us to
view and update a single user, our lens would look like this:

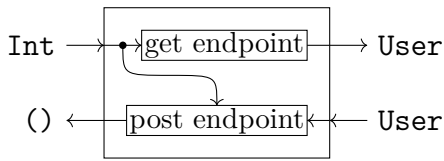
\begin{figure}[H]
    \centering
\begin{tikzpicture}

\draw (0.5,0) rectangle (3.5,2);
\draw[->] (0, 1.5) node [left] {\texttt{Int} } -- (0.5,1.5) ;
\draw[<-] (0, 0.5) node [left] {\texttt{()} } -- (0.5,0.5) ;
\draw[->] (3.5, 1.5) -- (4.0,1.5) node [right] {\texttt{User}};
\draw[<-] (3.5, 0.5) -- (4.0,0.5) node [right] {\texttt{User}};

    \draw (1,1.3) rectangle (3.2,1.7) ;
    \draw (2.1, 1.5) node {\small get endpoint} ;
    \draw [->] (0.5,1.5) -- (1,1.5) ;
    \draw (3.2,1.5) -- (3.5,1.5) ;
    \draw [->](0.75,1.5) node[circle,fill,inner sep=1pt] {} .. controls (0.75, 1.1) .. (1.375, 1.1) .. controls (2, 1.1) .. (2,0.7) ;
    \draw (0.8,0.3) rectangle (3.2,0.7) ;
    \draw (2, 0.5) node {\small post endpoint} ;
    \draw (0.5,0.5) -- (0.8,0.5);
    \draw [<-] (3.2,0.5) -- (3.5,0.5);

\end{tikzpicture}
\end{figure}

With this setup, creating a pair of endpoints exposing the \texttt{Address}
resource can be done by composing it with a lens $(User, User) \to (Address, Address)$:

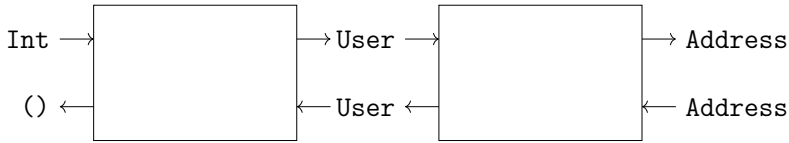
\begin{figure}[H]
    \centering
    \scalebox{0.9}{%
    \begin{tikzpicture}

    \draw (0.4,0) rectangle (3.4,2);
    \draw[->] (-0.1, 1.5) node [left] {\texttt{Int}} -- (0.4,1.5) ;
    \draw[<-] (-0.1, 0.5) node [left] {\texttt{()}} -- (0.4,0.5) ;
    \draw[->] (3.4, 1.5) -- (3.9,1.5) ;
    \draw[<-] (3.4, 0.5) -- (3.9,0.5) ;

    \draw (5.5,0) rectangle (8.5,2);
    \draw[->] (8.5, 1.5) -- (9,1.5) node [right] {\texttt{Address}};
    \draw[<-] (8.5, 0.5) -- (9,0.5) node [right] {\texttt{Address}};
    \draw[->] (5, 1.5) node [left] {\texttt{User}}  -- (5.5,1.5) ;
    \draw[<-] (5, 0.5) node [left] {\texttt{User}}  -- (5.5,0.5) ;

    \end{tikzpicture}
    }
\end{figure}

We observe that the uri of this lens is a type \texttt{Int}, we will develop the tools to write
uri as types in Section~\ref{uris-as-products}.

\hypertarget{endpoints-as-parametrised-lenses}{%
\subsection{Pairs of Endpoints as Parameteried lenses}
\label{pairs-of-endpoints-as-parametrised-lenses}}

We've explained how servers \emph{look} like lenses from the perspective
of the client but this does not explain how to \emph{implement}
servers. From the previous section, we know we could, in principle,
build endpoints out of existing ones by extending their boundaries
with lenses, but how do we implement the endpoints to begin with?

If we look at the pair of functions encoding the access to a
resource, we see that we are unable to implement our server:

\begin{Verbatim}[fontsize=\small]
getEndpoint : uri -> responseGET
postEndpoint : uri -> requestBody -> responsePOST
\end{Verbatim}

A \texttt{GET} endpoint cannot rely solely on the \texttt{URI} to
serve the relevant information to the client, it needs access to a
\emph{state}. Similarly, a \texttt{POST} endpoint needs to both access
the state \emph{and} modify it to correctly implement its
functionality. To take into account the state we need to access and
modify we are going to add a type parameter to our endpoints:

\begin{Verbatim}[fontsize=\small]
getEndpoint : (inputState, uri) -> responseGET
postEndpoint : (inputState, uri) -> requestBody
            -> (outputState, responsePOST)
\end{Verbatim}

In essence, if the client sees a resource as a plain lens
$(X, S) \to (Y, R)$, the server
sees it as a lens parametrised over the state $(P, Q)$ in which it operates:
$(P,Q) \times (X,S) \to (Y,R)$.
%
%
We thus end up with a complete definition of server:

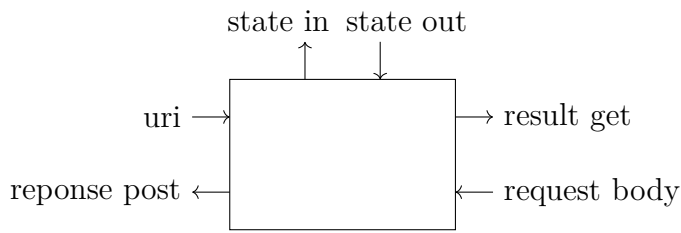
\begin{figure}[H]
    \centering
\begin{tikzpicture}
\draw (0.5,0) rectangle (3.5,2);
\draw[->] (0, 1.5) node [left] {uri } -- (0.5,1.5) ;
\draw[<-] (0, 0.5) node [left] {reponse post} -- (0.5,0.5) ;
\draw[->] (3.5, 1.5) -- (4.0,1.5) node [right] {result get};
\draw[<-] (3.5, 0.5) -- (4.0,0.5) node [right] {request body};

\draw[->] (1.5, 2) -- (1.5,2.5) node [above,xshift=-10] {state in};
\draw[<-] (2.5, 2) -- (2.5,2.5) node [above,xshift=10] {state out};
\end{tikzpicture}
\end{figure}

Before we move on to the implementation of our server, we need to explain how
this is a good enough representation of servers. First of all, parametrised
lenses capture resources and state updates using their top boundary. And using \emph{post-composition}
we can focus on deeper parts of a resource. Secondly, lawful lenses represent
idempotent resources that can be called with a \texttt{PUT} request for their updates.
Finally, unlawful lenses represent all other server endpoints that are not necessarily
resources but that we still need to represent.

But if post-composition represents focusing on a deeper part of the resource,
what do other lens operations such as  \emph{pre-composition} and \emph{parallel composition}
do for us? In what follows we explain how to use them for URI manipulation.

\hypertarget{uris-as-products}{%
\subsubsection{URIs as products}\label{uris-as-products}}

URIs represent paths to a resource, traditionally from a filesystem, but
with modern servers, they allow routing requests. Because
we use URIs to route requests, the chief property they have for us
is that they are \emph{parsable}. What is more, they are
\emph{structured} into path components, which can either be \emph{captures}
or \emph{path components}.

We decided to use right-nested products to represent URIs as types. This
way, \texttt{a * (b * (c * d))} represents the uri \texttt{a/b/c/d}.
With this representation, we can encode captures as
\emph{types} and path components as \emph{singleton}
types, such as:

\begin{Verbatim}[fontsize=\small]
data Str : String -> Type where
  MkS : (str : String) -> Str str
\end{Verbatim}

This data declaration ensures the type \texttt{Str "hello"} only
has a single inhabitant \texttt{MkS "hello"}.

We ensure that each path is parsable by requiring that each capture
is parsable. Singleton strings are trivially parsable using an equality
check.

\hypertarget{projections-as-path-extensions}{%
\subsubsection{Projections as path extensions}\label{projections-as-path-extensions}}

Using \emph{pre-composition} we can compose a lens to the left of
an endpoint that will perform some pre-processing on its left
boundary. Because we are interested in performing operations
such as declaring an endpoint as living under a higher directory
level, we want to model \emph{URI extension} as a lens.

\begin{figure}
\begin{tikzpicture}

\draw (0.5,0) rectangle (3.5,2);
\draw[->] (0, 1.5) node [left] {\texttt{/user/Int:id}} -- (0.5,1.5) ;
\draw[<-] (0, 0.5) node [left] {\texttt{()}} -- (0.5,0.5) ;
\draw[->] (3.5, 1.5) -- (4,1.5) ;
\draw[<-] (3.5, 0.5) -- (4,0.5) ;

    \draw (1.75,1.3) rectangle (2.25,1.7) ;
    \draw (2.0, 1.5) node {$\pi_2$} ;
    \draw[->] (2.25, 1.5) -- (3.5, 1.5);
    \draw[->] (0.5, 1.5) -- (1.75, 1.5);

    \draw[->] (6.5, 2) -- (6.5, 2.5) node (0)[xshift=-18, above] {\texttt{List User}};
    \draw[->] (7.5, 2) -- (7.5, 2.5) node (1)[xshift= 18, above] {\texttt{List User}};

\draw (5.5,0) rectangle (8.5,2);
\draw[->] (8.5, 1.5) -- (9,1.5) node [right] {\texttt{User}};
\draw[<-] (8.5, 0.5) -- (9,0.5) node [right] {\texttt{User}};
\draw[->] (5, 1.5) node [left] {\texttt{Int}}  -- (5.5,1.5) ;
\draw[<-] (5, 0.5) node [left,xshift=-4] {\texttt{()}}  -- (5.5,0.5) ;

    \draw (1.75,0.3) rectangle (2.25,0.7) ;
    \draw (2.0, 0.5) node {id} ;
    \draw[<-] (2.25, 0.5) -- (3.5, 0.5);
    \draw[<-] (0.5, 0.5) -- (1.75, 0.5);

\end{tikzpicture}
\caption{URI extension using pre-composition with an adapter}
\label{extended-uri-diagram}
\end{figure}
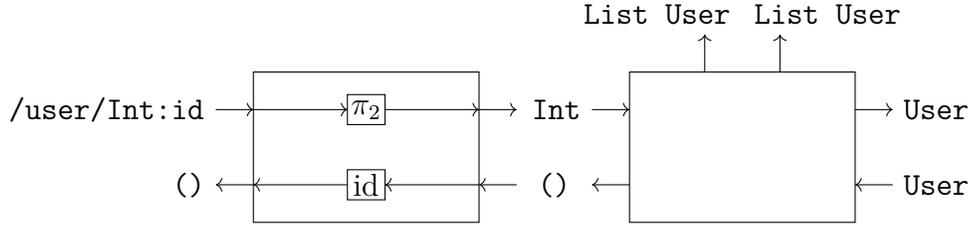

Thanks to our representation of URIs as products we can achieve this
by taking a URI and nesting it to the right of a product:
\begin{Verbatim}[fontsize=\small]
prependPath : String -> (uri : Type) -> Type
prependPath str path = Str str * path
\end{Verbatim}

Similarily, to add a capture to a URI we nest the existing path
to the right of the type of our capture:

\begin{Verbatim}[fontsize=\small]
-- Carrier is just a wrapper around types that are parsable
prependType : (ty : Type) -> Parsable ty => (uri : Type) -> Type
prependType ty path = Carrier ty * path
\end{Verbatim}

We implement the extension of paths with an adapter that has the
extended path as its left boundary and the projected path as its
right boundary. Figure \ref{extended-uri-diagram} shows a server
extended by such an adapter.

We abstract this operation with the operator \texttt{(/)}:

\begin{Verbatim}[fontsize=\small]
(/) : (str : String) -> Lens (X, S) (Y, R) ->
      Lens (Str str * X, S) (X, Y)
\end{Verbatim}
\hypertarget{captures-as-parallel-composition}{%
\subsubsection{Captures as parallel composition}\label{captures-as-parallel-composition}}

If we want to add a URI capture to an existing server we have to
use parallel composition on our endpoint with an identity lens that will
carry the capture to the lens that will make use of it.
In the following diagram, we make use of a \texttt{state} lens $(s)$
that exposes the state as a right boundary. We then use parallel
composition with an identity adapter that forwards the URI capture
to the right boundary of our \texttt{state} lens. Then we
\emph{post-compose} the server with a lens $(op)$ that performs \texttt{lookup}
and \texttt{insert} operations. Additionally, we \emph{pre-compose}
this server with an adapter $(p)$ which converts the left boundary into
a valid URI and a response body.

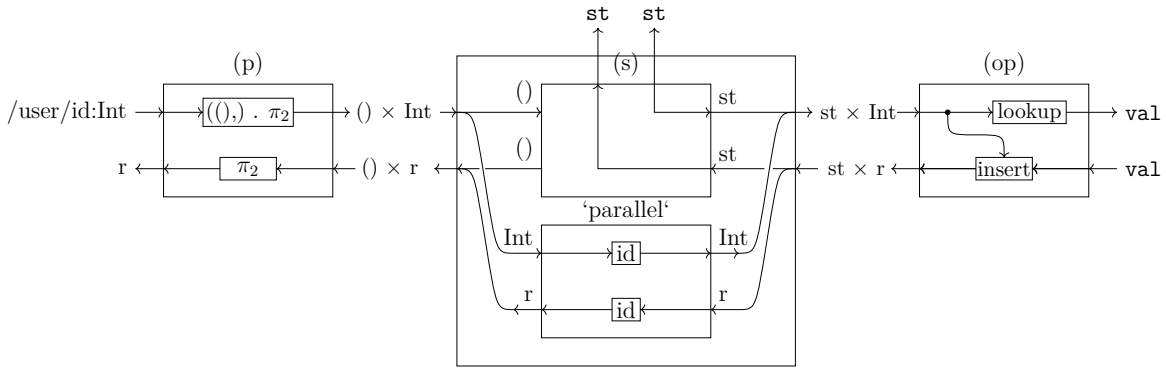
\scalebox{0.75}{%
\begin{tikzpicture}[every text node part/.style={align=center}]

\draw (0.5,0) rectangle (3.5,2);
\draw[->] (-1, 1.5)  -- (0.5,1.5) node [anchor=south east] {()} ;
\draw (-0.5, 0.5)  -- (0.5,0.5) node [anchor=south east] {()};
\draw[<-](-1, 0.5)  -- (-0.6,0.5) ;
\draw (3.5, 1.5) node [anchor=south west] {st} -- (5,1.5) ;
\draw[<-] (3.5, 0.5) node [anchor=south west] {st} -- (4.45,0.5) ;
\draw (4.6, 0.5)  -- (5.0,0.5) ;
    \draw[->] (2.5,2) -- (2.5,1.5) -- (3.5,1.5);
    \draw[<-] (1.5,2) -- (1.5,0.5) -- (3.5,0.5);

    \draw[->] (1.5, 2) -- (1.5, 3) node (0)[xshift=0, above] {\texttt{st}};
    \draw[->] (2.5, 2) -- (2.5, 3) node (1)[xshift=0, above] {\texttt{st}};
    \draw (2,2) node [above] {(s)} ;

\draw (8.7, 2) node [above] {(op)} ;
\draw (7.2,0) rectangle (10.2,2);
    \draw[->] (10.2, 1.5) -- (10.7,1.5) node [right] {\texttt{val}};
    \draw[<-] (10.2, 0.5) -- (10.7,0.5) node [right] {\texttt{val}};
    \draw[->] (6.8, 1.5) -- (7.2,1.5) ;
    \draw[<-] (6.7, 0.5) -- (7.2,0.5) ;

    \draw (8.5,1.3) rectangle (9.8,1.7) ;
    \draw (9.15, 1.5) node {\small lookup} ;
    \draw [->] (7.2,1.5) -- (8.5,1.5) ;
    \draw (9.8,1.5) -- (10.2,1.5) ;
    \draw [->](7.7,1.5) node[circle,fill,inner sep=1pt] {} .. controls (7.7, 1.1) .. (8.2, 1.1) .. controls (8.7, 1.1) .. (8.7,0.7) ;
    \draw (8.2,0.3) rectangle (9.2,0.7)  ;
    \draw (8.7,0.5) node  {\small insert};
    \draw [->](8.2,0.5) -- (7.2,0.5);
    \draw [<-] (9.2,0.5) -- (10.2,0.5);

\draw (0.5,-0.5) rectangle (3.5,-2.5);
    \draw[->] (3.5, -1) node [anchor=south west] {Int} -- (4,-1) ;
    \draw[<-] (3.5, -2) node [anchor=south west] {r} -- (4,-2) ;
    \draw[->] (0, -1)  -- (0.5,-1) node [anchor=south east] {Int};
    \draw[<-] (0, -2)  -- (0.5,-2) node [anchor=south east] {r};

    \draw (2, -0.25) node [] {`parallel`};

    \draw (1.75,-1.2) rectangle (2.25, -0.8) ;
    \draw (2, -1) node  {id} ;
    \draw[->] (2.25, -1) -- (3.5, -1.0);
    \draw[->] (0.5, -1) -- (1.75, -1.0);
    \draw (1.75,-2.2) rectangle (2.25,-1.8) ;
    \draw (2, -2) node {id} ;
    \draw[<-] (2.25, -2) -- (3.5, -2);
    \draw[<-] (0.5, -2) -- (1.75, -2);

\draw (-1, -3) rectangle (5, 2.5) ;
    \draw (4, -1) .. controls (4.3, -1) .. (4.5, 0.25) .. controls (4.7, 1.5) .. (5, 1.5);
    \draw (4, -2) .. controls (4.3, -2) .. (4.5, -0.75) .. controls (4.7, 0.5) .. (5, 0.5);
    \draw (0, -1) .. controls (-0.3, -1) .. (-0.5, 0.25) .. controls (-0.7, 1.5) .. (-1, 1.5);
    \draw (0, -2) .. controls (-0.3, -2) .. (-0.5, -0.75) .. controls (-0.7, 0.5) .. (-1, 0.5);
    \draw [->](5, 1.5) -- (5.3, 1.5) node [right] {\small st $\times$ Int};
    \draw [<-](5, 0.5) -- (5.4, 0.5) node [right,xshift=1] {\small st $\times$ r};
    \draw [->](-1.3, 1.5) node [left] {\small () $\times$ Int} -- (-1, 1.5);
    \draw [<-](-1.4, 0.5) node [left,xshift=-4] {\small () $\times$ r} -- (-1, 0.5);

    \draw (-4.7, 2) node [above] {(p)};
\draw (-6.2,0) rectangle (-3.2,2) ;
    \draw [<-] (-3.2, 0.5) -- (-2.8, 0.5) ;

    \draw [<-] (-6.7, 0.5) node [left] {r} -- (-6.2, 0.5) ;
    \draw [->] (-6.7, 1.5) node [left] {/user/id:Int} -- (-6.2, 1.5) ;

    \draw (-5.5,1.25) rectangle (-3.9,1.75)  ;
    \draw (-4.7, 1.5) node {\small ((),) . $\pi_2$} ;
    \draw [<-] (-2.9,1.5) -- (-3.9, 1.5) ;
    \draw [->] (-6.2,1.5) -- (-5.5, 1.5) ;

    \draw (-5.2,0.3) rectangle (-4.2,0.7)  ;
    \draw [->] (8.2,0.5) -- (7.2,0.5);
    \draw [<-] (9.2,0.5) -- (10.2,0.5);
    \draw (-4.7, 0.5) node {$\pi_2$} ;
    \draw [<-] (-4.2,0.5) -- (-3.2, 0.5) ;
    \draw [<-] (-6.2,0.5) -- (-5.2, 0.5) ;

\end{tikzpicture}
}

Because this careful threading of string diagrams is quite repetitive and
error-prone, we've abstracted this operation of carrying URI captures under
the operator \texttt{(:/)}.

\begin{Verbatim}[fontsize=\small]
-- Carrier carries around the evidence that `ty` is Parsable
(:/) : (ty : Type) -> Parsable ty => Lens (X, S) (Y, R) ->
       Lens (Carrier ty * X, S) (Carrier ty * Y, R)
\end{Verbatim}

Combined with our previous operator for path
extensions, this results in a familiar syntax to build
path components: \texttt{"user" / Int :/ "Todo" / …}

\hypertarget{server-implementation}{%
\subsection{Server implementation and limitations}\label{server-implementation}}

In this section, we explain at a high level how a collection of lenses can be translated
into a running server, and how parametrised lenses remain limited.

Because lenses represent resources and resources are pairs of endpoints,
Each lens generates a \GET and \POST endpoint which behave like data
accessors. That is, the forward part will implement the \GET
endpoint, and the backward part the \POST endpoint. Given a list of those lenses,
we can collect all the endpoints and their implementations and give them to a
\emph{server engine}. This engine will wait for requests, parse them using the parsers
from the list of lenses provided, and handle the request using the forward/backward
part (depending on if the request was \POST or \GET). Finally, the result will
be sent back to the client and its internal state will be updated.

This definition of servers is able to model URI parsing, \texttt{GET} and
\texttt{POST} endpoints, and create new endpoints by composing lenses with existing ones.
However, we observe two limitations:
\begin{itemize}
    \item All our lenses need to operate in the same state, otherwise we do not know
        how to update our state. But in principle, we would like to combine two lenses that
        operate in two different environments.
    \item We can combine our lenses in a list to build a server, but once in a list form, we cannot
        use lens operations to further extend and compose our server with other servers and endpoints.
\end{itemize}

While in practice this does not stop us from running servers, it does
fall short of our goal to find a data structure that explains server composition in
its entirely. Having to rely on an external program to perform routing and being unable
to combine servers of different states is inssuficient.

Both those limitations suggest that we are missing a composition operation that gives
us the choice of which lens to run. This is traditionally achieved with a
\emph{coproduct}: $A + B$.

Attempting to implement a coproduct on lenses will shed light on the next steps to take:

\begin{Verbatim}[fontsize=\small]
coproduct : Lens a b -> Lens x y -> Lens (a + x) (b + y)
coproduct (Lens get1 set1) (Lens get2 set2) =
  Lens (bimap get1 get2)
       (\(input : a.π1 + x.π1), (arg : b.π2 + y.π2) => ?what
\end{Verbatim}

It turns out that we cannot implement this operator on the backward part of the lens because
we only have access to functions \texttt{set1 : a.π1 -> b.π2 -> a.π2} and
\texttt{set2 : x.π1 -> y.π2 -> x.π2}. Unfortunately, the type signature we need to
implement allows for inputs of type \texttt{a.π1} and \texttt{y.π2} to be available
simultaneously, even if we do not have access to a function \texttt{a.π1 -> y.π2 -> a.π2 + x.π2}.

But if were able to constrain the type of \texttt{arg} to precisely correspond to the lens
taken by the input, then we could ensure that \texttt{arg} has type \texttt{b.π2} whenever \texttt{input} is
a \texttt{Left} value.

This property can only be achieved by introducing dependent types in our lenses. Which is
the topic of the next section.

	\hypertarget{dependent-parametrised-lenses}{
\section{Dependent (parametrised) lenses}\label{dependent-parametrised-lenses}}

Much like Section~\ref{polymorphic-parametrised-lenses}, we define lenses and their
parametrised counterparts, however, we are going to replace our
\emph{boundaries} with dependent boundaries. This extension of lenses toward dependent
lenses should provide us with the missing coproduct operator on lenses.

Previously our definition of lenses looked like this:

\begin{Verbatim}[fontsize=\small]
record Lens (l, r : Boundary) where
  constructor MkLens
  view : l.π1 -> r.π1
  update : l.π1 -> r.π2 -> l.π2
\end{Verbatim}

We replace our \texttt{Boundary} type by a dependent boundary \texttt{(x : Type, x -> Type)}
\begin{Verbatim}[fontsize=\small]
record DBoundary where
  π1 : Type
  π2 : π1 -> Type

public export
record DLens (l, r : DBoundary) where
  view : l.π1 -> r.π1
  update : (v : l.π1) -> r.π2 -> l.π2
\end{Verbatim}

This program does not typecheck because \texttt{l.π2} requires an argument of type \texttt{l.π1} and
similarily \texttt{r.π2} expects a value of type \texttt{r.π1}. We can easily find a value of type
\texttt{l.π1}, it's \texttt{v}, but we need a little bit more gymnastic to find a value of \texttt{r.π1}
and we do this by making a call to \texttt{view} in the type of \texttt{update}. Our corrected
definition looks like so:

\begin{Verbatim}[fontsize=\small]
record DLens (l, r : DBoundary) where
  view : l.π1 -> r.π1
  update : (v : l.π1) -> r.π2 (view v) -> l.π2 v
\end{Verbatim}

The \texttt{DLens} data type is also known in the litterature as \emph{container
morphism}\cite{abbott-etal-categories-containers, Abbott_Altenkirch_Ghani_McBride_2003}. With this in mind
we replace our boundaries by Containers and settle on this definition:

\begin{Verbatim}[fontsize=\small]
record Container where
  shape : Type
  position : shape -> Type

record DLens (l, r : Container) where
  constructor MkDLens
  view : l.shape -> r.shape
  update : (v : l.shape) -> r.position (view v) -> l.position v
\end{Verbatim}

The intuition of lenses as container morphisms suggests that, if we think of containers as type
descriptors and morphisms between them as maps between types, then lenses convert from one type
to the other and the implementation of the lens
performs the transformation between terms described by the containers.
However, the analogy breaks down
when we look at the forward part and the backward part of the lens as two endpoints modifying
different kinds of data. This curious disconnect will be revisited in future work (see Section~\ref{conclusion-future}).
For now, we are going to consider dependent lenses as lenses with additional structure, we start
by reiterating the operators we use on containers for dependent lens composition.

\subsection{Operations on Containers}

To combine our servers we need to understand how to combine their boundaries,
for this we quickly summarize the operations on containers as seen in
\cite{abbott-etal-categories-containers}

\subsubsection{Multiplication $(*)$}
This operator performs a product on the
shapes but a coproduct on the positions.

\begin{Verbatim}[fontsize=\small]
(*) : (c1, c2 : Container) -> Container
(*) c1 c2 = MkCont
    (c1.shp * c2.shp)
    (\x => c1.pos x.fst + c2.pos x.snd)
\end{Verbatim}
\subsubsection{Addition $(+)$} This operator performs a coproduct on
the shapes and, depending on which shape we are looking at, returns the position
associated with the first or second container. This represents a choice of containers.
\begin{Verbatim}[fontsize=\small]
(+) : (c1, c2 : Container) -> Container
(+) c1 c2 = MkCont
    (c1.shp + c2.shp)
    (either c1.pos c2.pos)
\end{Verbatim}
\subsubsection{Tensor} This operator performs a product on both the shapes
and positions. This represents a parallel composition of containers.
\begin{Verbatim}
tensor : (c1, c2 : Container) -> Container
tensor c1 c2 = MkCont
    (c1.shp * c2.shp)
    (\x : c1.shp * c2.shp =>
        c1.pos x.π1 * c2.pos x.π2)
\end{Verbatim}
\subsubsection{Const} This operator builds a container using the same type for
the shapes and the positions.
\begin{Verbatim}
Const : Type -> Container
Const ty = MkCont ty (const ty)
\end{Verbatim}

\subsection{Dependent Parametrised Lenses}

Because dependent lenses come equipped with the same composition and parallel
composition operations as plain lenses, we elude them and jump straight into
their parametrised variants, which we use for servers.

Just like in Section~\ref{parametrised-lenses}, we are going to extend our notion
of dependent lenses by adding a \emph{parameter} to our lenses, this parameter will
be the top boundary of our lenses which will represent the state in our server. We
define dependent parametrised lenses by using the tensor product to combine the
top boundary and the left boundary:

\begin{Verbatim}[fontsize=\small]
DPara : (l, p, r : Container) -> Type
DPara l p r = DLens (l `tensor` p) r
\end{Verbatim}

Just like before we unlock several operations on dependent parametrised lenses:

\subsubsection{Reparametrisation}. Just like plain parametrised lenses, we can modify
        the top boundary using a dependent lens.
\begin{Verbatim}[fontsize=\small]
reparam : DPara l p r -> DLens p' p -> DPara l p' r
reparam lens para = MkDLens
  (lens.view . mapSnd para.view)
  (\(x, y) => mapSnd (para.update y)
            . lens.update (x, para.view y))
\end{Verbatim}
\subsubsection{Sequential composition}. Just like with plain parametrised lenses, the top
        boundaries are tensored.
\begin{Verbatim}[fontsize=\small]
associate : a * (b * c) -> (a * b) * c

-- we use a different operator to differenciate with
-- composition of dependent lenses.
(|>) : DPara l p x -> DPara x q r ->
       DPara l (p `tensor` q) r
(|>) (MkDLens v1 u1) (MkDLens v2 u2) = MkDLens
  (v2 . mapFst v1 . associate)
  (\(x, (y, z)), arg => let
       { (v1, st1) = u2 (v1 (x, y), z) arg
       ; (v2, st2) = u1 (x, y) v1
       } in (v2, st2, st1))
\end{Verbatim}

\subsubsection{Pre-composition $(\lll)$ and Post-composition $(\ggg)$}
        Those operations
        are implemented in terms of composition and re-parametrisation.

\begin{Verbatim}[fontsize=\small]
-- Pre-composition
(<<<) : DLens l l' -> DPara l' p r -> DPara l p r

-- Post-composition
(>>>) : DPara l p r' -> DLens r' r ->  DPara l p r
\end{Verbatim}

\subsubsection{Parallel composition} Parallel composition is inherited from parallel composition
        of dependent lenses, composed with re-associating the left boundary.

\begin{Verbatim}[fontsize=\small]
reassoc : (a * x) * (b * y) -> (a * b) * (x * y)

parallel : DPara a p b -> DPara x q y ->
           DPara (a `tensor` x) (p `tensor` q) (b `tensor` y)
parallel l r = MkDLens reassoc (\((a, x), (b, y)) => reassoc)
            |> parallel l r -- This is the `parallel` operator
                            -- from DLens and |> is sequential
                            -- composition from DPara
\end{Verbatim}

\subsubsection{External choice $(+\&\&\&+)$} This is the `coproduct-like' operator we were looking for; it allows
        a client to choose which lens to run. Its implementation relies on
        the bifunctorial nature of \texttt{(+)} and \texttt{(*)}.

\begin{Verbatim}[fontsize=\small]
distributive : (s + s') * (p * p') -> (s * p) + (s' * p')

extChoice : DPara l p r -> DPara x q y ->
            DPara (l + x) (p * q) (r + y)
extChoice (MkDLens v1 u1) (MkDLens v2 u2) = MkDLens
  (bimap v1 v2 . distributive)
  (\case { ((Right v), (p, q)) => mapSnd Right . u2 (v, q)
         ; ((Left  v), (p, q)) => mapSnd Left  . u1 (v, p)})
\end{Verbatim}

        Parameters of both lenses need to be available in order to be ready to handle either input.
        We notice that, in the backward part, both \texttt{p} and \texttt{q} are available but
        we only use the one that corresponds to choice of the input.
        We start using parallax diagrams to demonstrate the choice of two lenses:

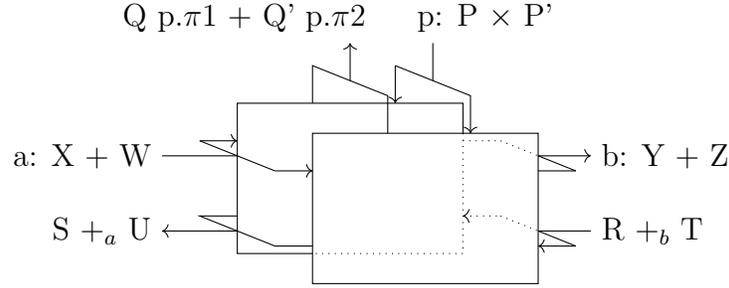
\begin{center}
\begin{tikzpicture}

\draw (1.5,-0.4) rectangle (4.5,1.6);
\draw (4.5, 1.1) -- (5,1.1) ;
\draw[<-] (4.5, 0.1) -- (5,0.1) ;
\draw[->] (1.0, 1.1) -- (1.5,1.1) ;
\draw (1.5,0) -- (0.5,0) -- (0.5,2) -- (3.5, 2) -- (3.5,1.6);
\draw[dotted] (3.5,1.6) -- (3.5,0) -- (1.5,0) ;
\draw[->] (0, 1.5) -- (0.5,1.5) ;
\draw[dotted] (3.5, 1.5) -- (4.0,1.5) ;
\draw[dotted, <-] (3.5, 0.5) -- (4.0,0.5) ;


\draw (1.5,2) -- (1.5,2.5) -- (2.5,2.1) -- (2.5,1.6) ;
\draw[<-] (2.6,2) -- (2.6,2.5);
\draw[->] (2.6,2.5) -- (3.6,2.1) -- (3.6,1.6) ;
    \draw[->] (2.0,2.3) -- (2.0, 2.8) node [above, xshift=10, anchor=south east] {Q p.π1 $+$ Q' p.π2};
\draw (3.1,2.3) -- (3.1, 2.8) node [above, xshift=-10, anchor=south west] {p: P $\times$ P'};

\draw[dotted] (4.0, 1.5) -- (4.5,1.3) ;
\draw (4.5, 1.3) -- (5.0,1.1) ;
\draw[->] (4.5, 1.3) -- (5.2,1.3) node [right] {b: Y $+$ Z};
\draw[dotted] (4.0, 0.5) -- (4.5,0.3) ;
\draw (4.5, 0.3) -- (5.0,0.1) ;
    \draw (4.5, 0.3) -- (5.2,0.3) node [right] {R $+_b$ T};
\draw (1.5,0.1) -- (1.0,0.1) -- (0,0.5) -- (0.5,0.5);
    \draw (0, 1.5) -- (1.0,1.1) ;

\draw (0.5, 1.3) -- (-0.5,1.3) node [left] {a: X $+$ W};
\draw[->]  (0.5, 0.3) -- (-0.5,0.3) node [left] {S $+_a$ U};

\end{tikzpicture}
\end{center}

$a +_i b$ is shorthand for \texttt{either a b i}.

Because this is the crux of our implementation of servers as lenses we take some time
to carefully explain how to read this result.

Looking at the diagram, an incoming request will pick which endpoint to call
by providing a value of type \texttt{X} or \texttt{W}. For the purposes of this explanation,
let us assume that the client calls a \GET endpoint with a uri \texttt{X}.
From this, the forward part of the lens in the background will be run and return
a value of type \texttt{Y}, which will be sent back as a response.

If the client calls the server with a \POST request, and a uri \texttt{X}, the
backward part will be run. But before it does, the type of the backward part needs
to be computed. That is because depending on which endpoints we call, we expect
to parse different request bodies. To find out the type of the request body, we
run the forward part with our value \texttt{a : X} to obtain a value \texttt{b : Left Y},
which we use to compute the type of the request body. Computing this type amounts
to evaluating the function \texttt{either R T b}, because we got a \texttt{Left Y}, the
request body is expected to be \texttt{R}. We attempt to parse the request body as
a value of type \texttt{R} and if successful, we finally run the backward part
with arguments \texttt{a} and request body of type \texttt{R}.
With the result of the backward part, we update the state with a value of type
\texttt{Q p.π1} and send a response back of type \texttt{S}.

To summarise, if the client provides an input of type \texttt{X} , the lens in the background
of the diagram is run, otherwise, the lens in the foreground is run.

\subsubsection{Clone choice $(\&\&\&)$} Clone choice is our final operator and is
    defined by reparametrisation using
    operators \texttt{dup : x -> x * x} and \texttt{dia : x + x -> x}.
    This operation allows the choice of two lenses to share the same parametr.
\begin{Verbatim}[fontsize=\small]
cloneChoice : DPara l p r -> DPara x p y ->
              DPara (l + x) p (r + y)
cloneChoice l1 l2 =
    reparam (extChoice l1 l2) (MkDLens dup (\_ => dia))
\end{Verbatim}

Just like external choice, clone choice allows the client to pick which
endpoints to call using either \texttt{X} or \texttt{W} and depending
on the choice, the corresponding lens is run. The difference is that
the state between those two lenses is now shared.

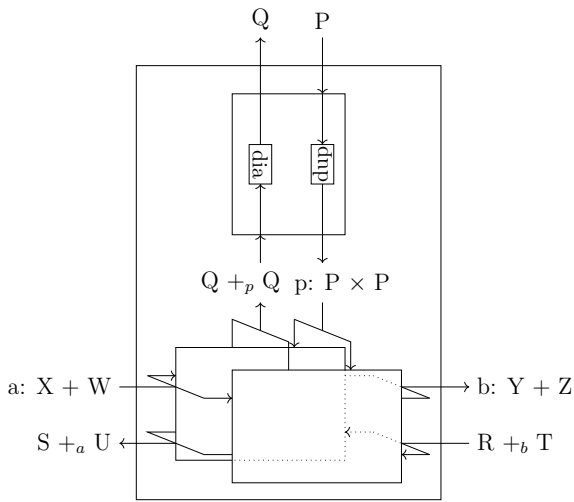
\begin{wrapfigure}{R}{0.5\textwidth}
\scalebox{0.75}{%
\begin{tikzpicture}

\draw (1.5,-0.4) rectangle (4.5,1.6);
\draw (4.5, 1.1) -- (5,1.1) ;
\draw[<-] (4.5, 0.1) -- (5,0.1) ;
\draw[->] (1.0, 1.1) -- (1.5,1.1) ;
\draw (1.5,0) -- (0.5,0) -- (0.5,2) -- (3.5, 2) -- (3.5,1.6);
\draw[dotted] (3.5,1.6) -- (3.5,0) -- (1.5,0) ;
\draw[->] (0, 1.5) -- (0.5,1.5) ;
\draw[dotted] (3.5, 1.5) -- (4.0,1.5) ;
\draw[dotted, <-] (3.5, 0.5) -- (4.0,0.5) ;


\draw (1.5,2) -- (1.5,2.5) -- (2.5,2.1) -- (2.5,1.6) ;
\draw[<-] (2.6,2) -- (2.6,2.5);
\draw[->] (2.6,2.5) -- (3.6,2.1) -- (3.6,1.6) ;
\draw[->] (2.0,2.3) -- (2.0, 2.8) node [above, xshift=-10] {Q $+_p$ Q};
\draw (3.1,2.3) -- (3.1, 2.8) node [above, xshift=10] {p: P $\times$ P};


\draw[<-] (3.1, 3.5) -- (3.1, 4.0) ;
\draw[->] (2.0, 3.5) -- (2.0, 4.0) ;
\draw (1.5, 4.0) rectangle (3.5, 6.5) ;
    \draw [->](2,4.0) -- (2, 4.9);
    \draw (1.8, 4.9) rectangle (2.2, 5.6) ;
    \draw (2, 5.25) node [rotate=-90] {\small dia} ;
    \draw (2, 5.6) -- (2, 6.5) ;
    \draw (3.1,4.0) -- (3.1, 4.9);
    \draw [<-](3.1,5.6) -- (3.1, 6.5);
    \draw (2.9, 4.9) rectangle (3.3, 5.6) ;
    \draw (3.1, 5.25) node [rotate=90] {\small dup} ;

\draw[<-] (3.1, 6.5) -- (3.1, 7.5) node [anchor=south] {P};
\draw[->] (2.0, 6.5) -- (2.0, 7.5) node [anchor=south] {Q};

\draw[dotted] (4.0, 1.5) -- (4.5,1.3) ;
\draw (4.5, 1.3) -- (5.0,1.1) ;
\draw[->] (4.5, 1.3) -- (5.7,1.3) node [right] {b: Y $+$ Z};
\draw[dotted] (4.0, 0.5) -- (4.5,0.3) ;
\draw (4.5, 0.3) -- (5.0,0.1) ;
    \draw (4.5, 0.3) -- (5.7,0.3) node [right] {R $+_b$ T} ;
\draw (1.5,0.1) -- (1.0,0.1) -- (0,0.5) -- (0.5,0.5);

\draw (0, 1.5) -- (1.0,1.1) ;
\draw (0.5, 1.3) -- (-0.5,1.3)node [left] {a: X $+$ W};
\draw[->] (0.5, 0.3) -- (-0.5,0.3) node [left] {S $+_a$ U};

    \draw (-0.2, -0.7) rectangle  (5.2, 7);

\end{tikzpicture}
}
\caption{Clone choice operator implemented as a reparametrisation of external choice}
\end{wrapfigure}

	\hypertarget{server-as-dependent-lenses}{%
\section{Servers as dependent parametrised lenses}\label{server-as-dependent-lenses}}

Previously we saw that parametrised lenses were not powerful
enough to build entire servers because they lacked a coproduct,
since dependent parametrised lenses provide this coproduct we are
updating our definitions to make use of it.

In what follows, because there is no meaningful distinction
between endpoints, resources, and servers, we are going to refer to
all of them as ``servers''. Similarly, because each dependent parametrised
lens corresponds to a server, we are going to refer to them as
\texttt{Server} in code samples rather than \texttt{DPara}.

\hypertarget{ext-choice-composition}{%
    \subsection{External choice for state composition}\label{ext-choice-composition}}

The first operator we will use is \texttt{extChoice}, which allows
combining two resources to form a server.

\begin{figure}
\begin{tikzpicture}[every text node part/.style={align=center}]

\draw (1.5,-0.4) rectangle (4.5,1.6);
\draw (4.5, 1.1) -- (5,1.1) ;
\draw[<-] (4.5, 0.1) -- (5,0.1) ;
\draw[->] (1.0, 1.1) -- (1.5,1.1) ;
\draw (1.5,0) -- (0.5,0) -- (0.5,2) -- (3.5, 2) -- (3.5,1.6);
\draw[dotted] (3.5,1.6) -- (3.5,0) -- (1.5,0) ;
\draw[->] (0, 1.5) -- (0.5,1.5) ;
\draw[dotted,->] (3.5, 1.5) -- (4.0,1.5) ;
\draw[dotted, <-] (3.5, 0.5) -- (4.0,0.5) ;


\draw (1.5,2) -- (1.5,2.5) -- (2.5,2.1) -- (2.5,1.6) ;
\draw[<-] (2.6,2) -- (2.6,2.5);
\draw[->] (2.6,2.5) -- (3.6,2.1) -- (3.6,1.6) ;
\draw[->] (2.0,2.3) -- (2.0, 2.8);
\draw (3.1,2.3) -- (3.1, 2.8) ;
    \draw (2.55, 2.8) node [above] {Const (Dict ID User) \\
                                  * Const (Dict ID Todo)};

\draw[dotted] (4.0, 1.5) -- (4.5,1.3) ;
\draw (4.5, 1.3) -- (5.0,1.1) ;
    \draw[->] (4.5, 1.3) -- (5.2,1.3) node [right] {b: User + Todo};
\draw[dotted] (4.0, 0.5) -- (4.5,0.3) ;
\draw (4.5, 0.3) -- (5.0,0.1) ;
    \draw (4.5, 0.3) -- (5.2,0.3) node [right] {User $+_b$ Todo} ;
\draw (1.5,0.1) -- (1.0,0.1) -- (0,0.5) -- (0.5,0.5);

\draw (0, 1.5) -- (1.0,1.1) ;
    \draw (0.5, 1.3) -- (-0.5,1.3) node [left] {a: /user/id:Int\\+ /todo/id:Int};
\draw[->] (0.5, 0.3) -- (-0.5,0.3) node [left] {() $+_a$ ()};

\end{tikzpicture}
\caption{The external choice of two servers}
\label{ext-choice-servers}
\end{figure}
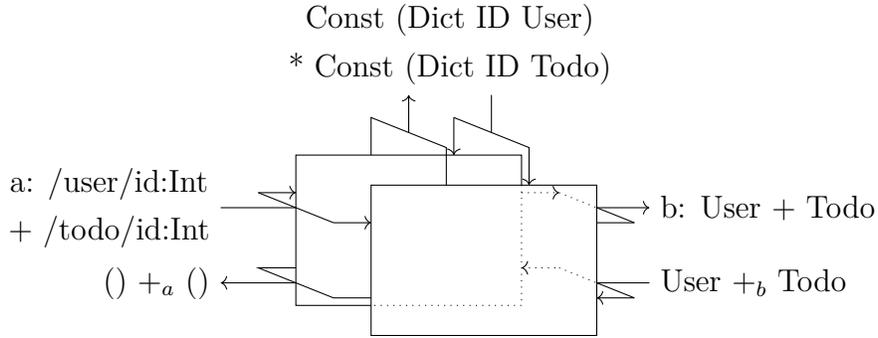

The external choice perfectly captures combining two endpoints
into a server by performing a co-product on the left boundary
of the lens, giving the client the choice of which
endpoint to call. The states are combined with a product to
ensure they both are available regardless of which endpoint
is called, this can be seen in the implementation of the backward
part of the lens:

\begin{Verbatim}[fontsize=\small]
  -- p and q are both available because we don't know
  -- if we will receive a `Left v` or a `Right v`
  (\case (Right v, (p, q)) => mapSnd Right . u2 (v, q)
         (Left  v, (p, q)) => mapSnd Left  . u1 (v, p))
\end{Verbatim}

As a concrete
example, Figure~\ref{ext-choice-servers} shows that
if we have a server with state \texttt{Dict ID User} and another
one with state \texttt{Dict ID Todo}, combining both results in
a server that works with state \texttt{Dict ID User * Dict ID Todo} \\

As a welcomed side effect, this buys us the ability to run each of those
servers concurrently since the states will not share any data.

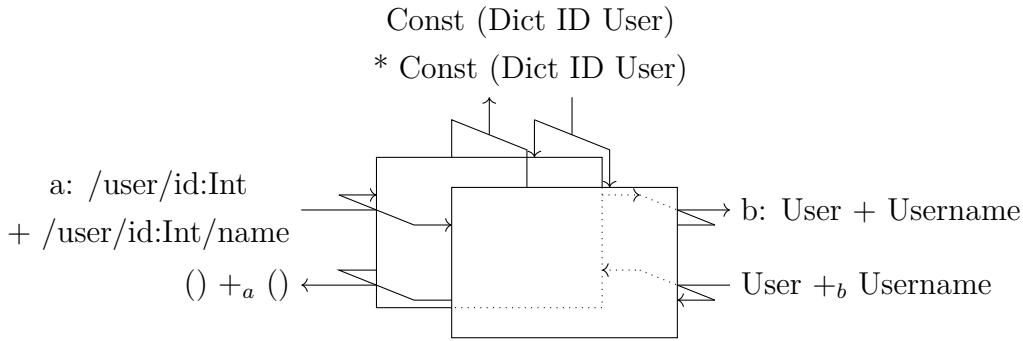
\begin{figure}
\begin{center}
\begin{tikzpicture}[every text node part/.style={align=center}]

\draw (1.5,-0.4) rectangle (4.5,1.6);
\draw (4.5, 1.1) -- (5,1.1) ;
\draw[<-] (4.5, 0.1) -- (5,0.1) ;
\draw[->] (1.0, 1.1) -- (1.5,1.1) ;
\draw (1.5,0) -- (0.5,0) -- (0.5,2) -- (3.5, 2) -- (3.5,1.6);
\draw[dotted] (3.5,1.6) -- (3.5,0) -- (1.5,0) ;
\draw[->] (0, 1.5) -- (0.5,1.5) ;
\draw[dotted,->] (3.5, 1.5) -- (4.0,1.5) ;
\draw[dotted, <-] (3.5, 0.5) -- (4.0,0.5) ;


\draw (1.5,2) -- (1.5,2.5) -- (2.5,2.1) -- (2.5,1.6) ;
\draw[<-] (2.6,2) -- (2.6,2.5);
\draw[->] (2.6,2.5) -- (3.6,2.1) -- (3.6,1.6) ;
\draw[->] (2.0,2.3) -- (2.0, 2.8);
\draw (3.1,2.3) -- (3.1, 2.8) ;
    \draw (2.55, 2.8) node [above] {Const (Dict ID User) \\
                                  * Const (Dict ID User)};

\draw[dotted] (4.0, 1.5) -- (4.5,1.3) ;
\draw (4.5, 1.3) -- (5.0,1.1) ;
    \draw[->] (4.5, 1.3) -- (5.2,1.3) node [right] {b: User + Username};
\draw[dotted] (4.0, 0.5) -- (4.5,0.3) ;
\draw (4.5, 0.3) -- (5.0,0.1) ;
    \draw (4.5, 0.3) -- (5.2,0.3) node [right] {User $+_b$ Username} ;
\draw (1.5,0.1) -- (1.0,0.1) -- (0,0.5) -- (0.5,0.5);

\draw (0, 1.5) -- (1.0,1.1) ;
    \draw (0.5, 1.3) -- (-0.5,1.3) node [left] {a: /user/id:Int\\+ /user/id:Int/name};
\draw[->] (0.5, 0.3) -- (-0.5,0.3) node [left] {() $+_a$ ()};

\end{tikzpicture}
\end{center}
\caption{The external choice of two servers with duplicated state}
    \label{duplicated-state}
\end{figure}

However, something is not right, if we combine two servers that
work in the same state, the state parameter is duplicated
after external choice, as you see in Figure~\ref{duplicated-state}.
Performing a state update through
the \POST endpoint of one server will not reflect the change in
the other server. To correct this, we will use \texttt{cloneChoice}.

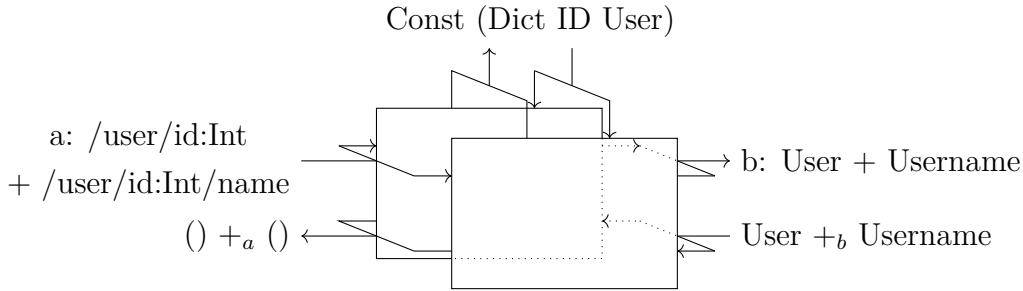
\begin{figure}
\begin{center}
\begin{tikzpicture}[every text node part/.style={align=center}]

\draw (1.5,-0.4) rectangle (4.5,1.6);
\draw (4.5, 1.1) -- (5,1.1) ;
\draw[<-] (4.5, 0.1) -- (5,0.1) ;
\draw[->] (1.0, 1.1) -- (1.5,1.1) ;
\draw (1.5,0) -- (0.5,0) -- (0.5,2) -- (3.5, 2) -- (3.5,1.6);
\draw[dotted] (3.5,1.6) -- (3.5,0) -- (1.5,0) ;
\draw[->] (0, 1.5) -- (0.5,1.5) ;
\draw[dotted,->] (3.5, 1.5) -- (4.0,1.5) ;
\draw[dotted, <-] (3.5, 0.5) -- (4.0,0.5) ;


\draw (1.5,2) -- (1.5,2.5) -- (2.5,2.1) -- (2.5,1.6) ;
\draw[<-] (2.6,2) -- (2.6,2.5);
\draw[->] (2.6,2.5) -- (3.6,2.1) -- (3.6,1.6) ;
\draw[->] (2.0,2.3) -- (2.0, 2.8);
\draw (3.1,2.3) -- (3.1, 2.8) ;
    \draw (2.55, 2.8) node [above] {Const (Dict ID User)};

\draw[dotted] (4.0, 1.5) -- (4.5,1.3) ;
\draw (4.5, 1.3) -- (5.0,1.1) ;
    \draw[->] (4.5, 1.3) -- (5.2,1.3) node [right] {b: User + Username};
\draw[dotted] (4.0, 0.5) -- (4.5,0.3) ;
\draw (4.5, 0.3) -- (5.0,0.1) ;
    \draw (4.5, 0.3) -- (5.2,0.3) node [right] {User $+_b$ Username} ;
\draw (1.5,0.1) -- (1.0,0.1) -- (0,0.5) -- (0.5,0.5);

\draw (0, 1.5) -- (1.0,1.1) ;
    \draw (0.5, 1.3) -- (-0.5,1.3) node [left] {a: /user/id:Int\\+ /user/id:Int/name};
\draw[->] (0.5, 0.3) -- (-0.5,0.3) node [left] {() $+_a$ ()};

\end{tikzpicture}
\end{center}
\caption{cloneChoice of two servers}
    \label{clone-choice-servers}
\end{figure}

\hypertarget{clone-choice-composition}{%
\subsection{Clone choice for endpoint composition}\label{clone-choice-composition}}

\texttt{cloneChoice}
combines two servers that operate in the same environment. This
will fix our issue with \texttt{extChoice} which duplicates the
state when combining two servers with the same state. With
\texttt{cloneChoice} the state is now shared between the two endpoints
in such a way that if a change is performed on one of the states,
it will be reflected in the state of the other server,
Figure~\ref{clone-choice-servers} illustrates the result of \texttt{cloneChoice}
on two servers sharing the same state.

\hypertarget{request-routing}{%
\subsection{Request routing}\label{request-routing}}

Previously in Section~\ref{server-implementation}, we had to rely on
an external mechanism, our list of request handlers, to route
our incoming requests to the
corresponding lens. Now that we can implement our entire server as a
single lens, we can rely on the parsing function of its input boundary
to perform routing.

We saw in Section~\ref{pairs-of-endpoints-as-parametrised-lenses} how to use composition and tensor for path
extension and captures. Those operations perform a coproduct and a product on
the input boundaries, and because our servers require their input
boundary to be parsable, the resulting boundary will remain parsable.

We use parser combinators \cite{parser_combinators} which are closed
over products and coproducts, through functions typically called
\texttt{sequence} and \texttt{alternative} respectfully:

\begin{Verbatim}[fontsize=\small]
-- run the left parser, if it fails, run the right parser
alternative : Parser a -> Parser b -> Parser (a + b)

-- run the first parser first, and follow it up by the second parser
sequence : Parser a -> Parser b -> Parser (a * b)
\end{Verbatim}

Because we use an \emph{interface}\footnote{Interfaces are similar to
typeclasses in haskell.} to enforce the parsable constraint
of the left boundary we can automatically derive implementation for
products and coproducts of parsers:

\begin{Verbatim}[fontsize=\small]
Parsable a => Parsable b => Parsable (a * b) where
  parse = sequence (parse {t=a}) (parser {=b})

Parsable a => Parsable b => Parsable (a + b) where
  parse = alternative (parse {t=a}) (parse {t=b})
\end{Verbatim}

We use a similar technique to enforce \emph{serialisation} constraints
on the response types. This way, any composition of two valid servers
with external choice or parallel composition results in a valid server.

\hypertarget{state-management}{%
\subsection{State management}\label{state-management}}

Previously, we were limited to servers whose resources all worked in the
same state. We have lifted this limitation by using containers
as boundaries. This upgrade comes with a caveat: The
engine performing state updates for the server is now unable to override
its previous state with the new state. This is because given a top boundary
\texttt{st}, the engine that runs the server expects an initial state of type
\texttt{p : st.π1}, but the result of running the backward part of our server lens
results in a value of type \texttt{st.π2 p}.
We can clearly see this inconsistency in the type of the backward part:

\begin{Verbatim}[fontsize=\small]
update : (v : p.π1 * l.π1) -> r.π2 (view v) -> p.π2 v.π1 * l.π2 v.π2
\end{Verbatim}

This is problematic for our state update because
when we update our internal state we perform a side effect of the sort:

\begin{Verbatim}[fontsize=\small]
-- pseudo code
handleRequest : Request -> IO ()
handleRequest req =
    do currentState <- get state
       (newState, result) <- handleRequest req currentState
       put newState -- update the state here
       response result
\end{Verbatim}

We overcome this issue by requiring every top boundary of a server to
carry an action of the shapes on the positions of our container.
We write this in terms of an interface on containers:

\begin{Verbatim}[fontsize=\small]
interface Action (c : Container) where
  act : (v : c.shape) -> c.positions v -> c.shape
\end{Verbatim}

Just like with parsing we provide implementations for products,
coproducts, and tensor which will ensure closure over the
action with regard to server composition:

\begin{Verbatim}[fontsize=\small]
interface Action (c : Container) where
  act : (s : c.shapes) -> c.positions s -> c.shapes

Action c1 => Action c2 => Action (c1 `tensor` c2) where
  act (s1, s2) (p1, p2) = (act s1 p1, act s2 p2)

Action c1 => Action c2 => Action (c1 + c2) where
  act (Left shape) p = Left (act shape p)
  act (Right shape) p = Right (act shape p)

Action c1 => Action c2 => Action (c1 * c2) where
  act (s1, s2) (Left x) = (act s1 x, s2)
  act (s1, s2) (Right x) = (s1, act s2 x)
\end{Verbatim}

Using our action we can patch up our previous
implementation to make it typecheck.

\begin{Verbatim}[fontsize=\small]
-- pseudo code
handleRequest : Request -> IO ()
handleRequest req = do
       overallState <- get state
       (newState, result) <- handleRequest req overallState
       put (act overallState newState) -- update the state here
       response result
\end{Verbatim}

We can think of the action as interpreting the \emph{diff}
of a substate in the overall state.

\hypertarget{implementation-as-server}{%
\subsection{Implementing a dependent lens as a server}\label{implementation-as-server}}

\begin{figure}
\begin{tikzpicture}[every text node part/.style={align=center}]
\draw (0.5,0) rectangle (3.5,2);
\draw[->] (0, 1.5) node [left,yshift=5] {x: coproduct of\\accepted URIs } -- (0.5,1.5) ;
\draw[<-] (0, 0.5) node [left,yshift=-5] {coproduct of \POST responses\\depending on $x$} -- (0.5,0.5) ;
\draw[->] (3.5, 1.5) -- (4.0,1.5) node [right,yshift=5] {r: coproduct of \GET\\responses computed from $x$};
\draw[<-] (3.5, 0.5) -- (4.0,0.5) node [right,yshift=-5] {coproduct of request body\\depending on $r$};

\draw[->] (1.5, 2) -- (1.5,2.5) ;
\draw[<-] (2.5, 2) -- (2.5,2.5) node [above,xshift=-15] {Product of states};
\end{tikzpicture}
    \caption{a Dependent parametrised lens as a server}
    \label{dep-para-server}
\end{figure}
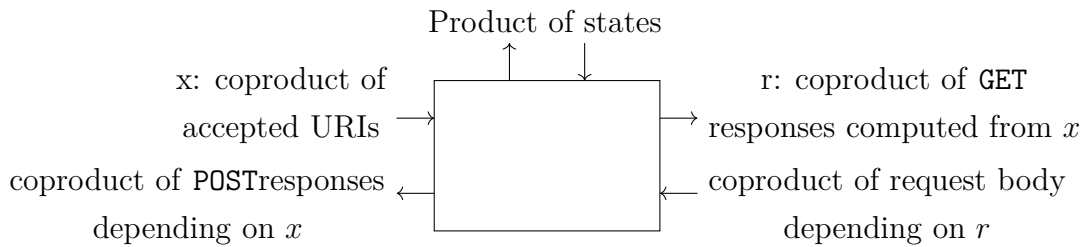

The definitions we gave culuminate in this section
where we explain how to convert a single lens to a
fully functional webserver. To summarise, a server is any dependent
parametrised lens $(P, Q)\times(X, S) \to (Y, R)$ such that
$X$ and $R$ are \texttt{Parsable}, $S$ and $Y$ are \texttt{Serialisable}
and $P$ and $Q$ have an \texttt{Action}. In Figure~\ref{dep-para-server},
we update the diagram in Section~\ref{resources-as-parametrised-lenses} to capture the fact that each
boundary depends on which lens was picked by the client when calling the
server.

From this, we can re-use the same architecture from
Section~\ref{resources-as-parametrised-lenses} but we only
need to generate 2 request handlers, one for \POST and one for
\GET requests, the rest is managed by the implementation of the lens.

In the following program, we achieved our goal to represent
a server as a single lens. The type informs us of the conditions
under which this abstraction makes sense.

\begin{Verbatim}[fontsize=\small, samepage=true]
-- `ISerialisable` and `IParsable` are dependent version of
-- `Parsable` and `Serialisable`
toHandler' : Parsable l.shape => IParsable r.position =>
             ISerialisable l.position => Serialisable r.shape =>
             Action p => Default p.shape =>
             Server l p r -> IO ()
toHandler' (MkDLens view update) =
   runServer (generateHandlers view update) def
\end{Verbatim}

We see that every lens gives rise to a server as long as its top left
and bottom right boundary are \texttt{Parsable}, and it's top right and
bottom left boundary are \texttt{Serialisable}. Additionally, the state
must have an accompanying \texttt{Action} and the product of states must
have a \texttt{Default} value which acts as the initial state.
\texttt{runServer} is our \emph{engine} that waits for requests and
provides responses.

	\hypertarget{the-recombine-library}{%
    \section{The \textsc{recombine} library}\label{the-recombine-library}}

The previous constructions have been implemented in Idris in a library
we call \textsc{recombine}. This library allows you to write lenses using
a small DSL and instanciate them into a running server.
Using the building blocks we presented earlier
we can write, extend and abstract over entire servers. To demonstrate
the library we showcase 3 examples of servers with varying levels of
complexity:
\begin{itemize}
    \item A calculator.
    \item An IoT server that controls devices.
    \item A todo App.
\end{itemize}
Those examples will culminate in a combined server that hosts
all three servers at once.

\hypertarget{calculator}{%
\subsection{Calculator}\label{calculator}}

Our calculator API has one endpoint per operation, each operation takes
two arguments and returns a result without accessing or modifying any
state. This is a simple example of a pure server running without state.
Here is the API we aim to implement:

\begin{verbatim}
/add/Int:n1/Int:n2
/sub/Int:n1/Int:n2
/mul/Int:n1/Int:n2
/div/Int:n1/Int:n2
\end{verbatim}

Using our DSL we can start with the \texttt{add} endpoint:

\begin{lstlisting}
addition : Server (MkCont (Int * Int) unit)
                  (Const ())
                  (MkCont Int unit)
addition = GetLens add
\end{lstlisting}

This endpoint accepts two integer numbers and returns another \texttt{Int}. It
is implemented using the smart constructor \texttt{GetLens} which only
populates the forward part and uses an identity as the backward part. We
extend this server on its left boundary to have it live
under the \texttt{/add} prefix:

\begin{lstlisting}
addEndpoint : Server (MkCont (Str "add" * Int * Int) unit)
                     (Const ())
                     (MkCont Int unit)
addEndpoint = "add" / addition
\end{lstlisting}
We define \texttt{sub}, \texttt{mul}, \texttt{div} similarity so we end
up with the following declarations:

\begin{lstlisting}
subEndpoint : Server (Str "sub" * Int * Int, unit) ? ?
mulEndpoint : Server (Str "mul" * Int * Int, unit) ? ?
divEndpoint : Server (Str "div" * Int * Int, unit) ? ?
\end{lstlisting}

We use \texttt{?} (question mark) in the type to indicate that we let Idris
infer the type. We leave them off except for the first one to document how the
type signature carries the API information.
We build the server by using the \texttt{cloneChoice} operator \texttt{(\&\&\&)}:

\begin{lstlisting}
calculator : Server (Str "add" * Int * Int
                   + Str "sub" * Int * Int
                   + Str "mul" * Int * Int
                   + Str "div" * Int * Int) ? ?
calculator = addEndpoint &&& subEndpoint
         &&& mulEndpoint &&& divEndpoint
\end{lstlisting}

One of the benefits of this approach is that the entire API of the server is
visible \emph{in the type} of the server and is computed along with its
implementation. \hypertarget{iot-server}{%
\subsection{IOT server}\label{iot-server}}

This server aims to showcase the ability of the framework to be
extended with lenses that focus on deeper parts of a resource. We exemplify
this by means of a home
automation server whose state is the product of booleans which represent
the state of various devices:
\(bool \times (bool \times bool )\).

As a product of types, we can use lenses
$\text{fst : } ((a, b), (a, b)) \to (a, a)$ and
$\text{snd : } ((a, b), (a, b)) \to (b, b)$  to access and
update each element. Before we start, here is the API we want to
obtain:

\begin{Verbatim}[fontsize=\small]
GET  /boiler -> Bool
POST /boiler {Bool} → Bool
GET  /lights/1 → Bool
POST /lights/1 {Bool} → Bool
GET  /lights/1 → Bool
POST /lights/2 {Boot} → Bool
\end{Verbatim}

The type of the body expected for \POST requests is written in curly braces.

The first step is to declare an endpoint that views and updates the
state:

\begin{lstlisting}
stateEndpoint : Server (Const ()) (Const HomeState) (Const HomeState)
stateEndpoint = State HomeState
\end{lstlisting}

In order to access and modify the boiler, we need to post-compose a lens
that focuses on the first component of the product. We also need
to pre-compose a lens that parses the \texttt{/boiler} URI. We combine
both operations in this definition:

\begin{lstlisting}
boilerEndpoint : Server (MkCont (Str "boiler") unit)
                        (Const HomeState)
                        (Const Bool)
boilerEndpoint = "boiler" / stateEndpoint >>> fst
\end{lstlisting}

We explicitly wrote the type signature to demonstrate how,
for resources, the boundaries are \texttt{Const} containers.

To implement the endpoints for lights we create a parent endpoint
that exposes the lights, and we post-compose it with \texttt{fst}
and \texttt{snd} lenses to give access to each light individually.

\begin{lstlisting}
lights : Server ? ? ?
lights = stateEndpoint >>> snd

iotServer = Server ? ? ?
iotServer = "boiler" / stateEndpoint >>> fst
        &&& "lights" / ("1" / lights >>> fst
                    &&& "2" / lights >>> snd)
\end{lstlisting}

We leave off the types completely in this example, this show how the framework
gives the choice of either declaring the API in advance
and let the types guide the implementation, or write the implementation and
check the API is suitable after typechecking.

\hypertarget{todo-app}{%
\subsection{Todo app}\label{todo-app}}

A Todo App is a perfect example that demonstrates how to implement a
real server with practical functionality. It shows how abandoning the lens
laws still results in a practical and composable server.
Here is the API we set out to write:

\begin{Verbatim}[fontsize=\small]
GET  /all/userId:Nat -> List Todo
POST /add/userId:Nat {Todo} -> ()
\end{Verbatim}

We implement this server by using \texttt{GetLens} defined previously
and \texttt{PostLens}, its \POST counterpart.

\begin{Verbatim}[fontsize=\small]
getTodos : Server (MkCont ? (const ()))
                  (Const ServerState)
                  (MkCont (List Todo) (const ()))
getTodos = "all" / Nat :/
    (GetLens (\(st, userId) =>
        fromMaybe [] . lookup st userId))

postTodo : Server ? (Const ServerState) ?
postTodo = "add" / Nat :/
    (PostLens (\(st, userId), todo =>
        update userId (todo ::) st))

-- The server is the composition of the previous two endpoints
todoServer : Server ? (Const ServerState) ?
todoServer = getTodos &&& postTodo
\end{Verbatim}

\hypertarget{combining-multiple-servers}{%
\subsection{Combining multiple servers}\label{combining-multiple-servers}}

Because the external choice of servers results in a valid server, we expect
to be able to combine multiple servers of drastically
different nature together. Here is how we combine the three servers we just
defined into one:

\begin{lstlisting}
combinedServer: Server ? (Const () *
                          Const ServerState *
                          Const HomeState) ?
combinedServer = "todo" / todoServer
                 +&&&+
                 "calculator" / calculator
                 +&&&+
                 "iot" / iotServer
\end{lstlisting}

We explicitely write down the type of the state, to demonstrate that
it is correctly composed using \texttt{(*)} on containers.

\hypertarget{conclusion-future}{%
\section{Conclusion and future work}\label{conclusion-future}}

We found a similarity in the way we represent data access using
lenses and the way servers expose their API. This approach
motivated the use of parametrised lenses, dependent lenses, and
dependent parametrised lenses, which turned out to be the
missing abstraction to explain server composition that handles
state composition, routing, and nested state updates.

We are going to conclude on a note about dependent types.
While lenses from Section~\ref{parametrised-lenses}
are good enough to describe single resources, only
\emph{dependent lenses} provides a complete feature set for
a server library.
Dependent types help tremendously in the
user experience of writing servers. Indeed the documentation
of the API is readily available while writing the server. And
because the boundaries can be inferred from the implementation,
the programmer has the choice of either writing the server first
and then check its API, or write the API first and then implement
a program that serves it. One could conceivably expect a mechanism
such as \emph{type providers} to read a specification file, such
as OpenAPI, and generate an API \emph{type} to implement.

Despite being successful in describing servers and implementing
them, this work prompts a lot of questions that we leave for
future work, we summarize them here.

While we can derive server APIs and implementation from dependent
lenses, and themselves are container morphisms, we should
expect more interaction with the different semantics given by the
$\mathrm{Poly}$ and $\mathrm{Cont}$ interpretation of our lens boundaries. For example,
we expect to be able to express aspects of server
communication using dependent lenses and interpret them
as a protocol that handles errors if the protocol is not respected.
This research area should prove fruitful when combined with
indexed containers \cite{Altenkirch_Ghani_Hancock_Mcbride_Morris_2015}
to represent legal transitions in
stateful protocols. This approach could tell us something about
implementing stateful servers and link \emph{session types} with
our description of server based on container morphisms.
In practice, this would enable the same framework to be reused
with a different \emph{engine} than HTTP and find itself useful for
describing UDP servers or Bluetooth communications.

Another unanswered question is the interpretation of dependent
parametrised morphisms as an interactive system the categorical
cybernetics way \cite{towards_foundations_categorical_cybernetics}.
In it, parametrised lenses represent bidirectional systems with a nesting, interacting controller acting as on the system through the `top boundary'.
In that interpretation, we see the parameter as
\emph{guiding} the system after it's been presented with an input, and \emph{observing} the outcome after a response has been computed.
There is an analogy between a server's routing mechanism guiding a
request toward its handler and an agent being guided through
the steps of a game. In future work, we plan to merge those two ideas,
parameters as performing routing, and polynomials as interactive
systems better describe the dynamic nature of servers.
In particular, we expect to even better model the interaction between routing and serving content from a storage, making it even more compositional and principled than proposed here.
The rough idea is to follow \cite{capucci2021translating} and thinking of URIs as `strategies' for parametrised data accessors on a stored state, thereby flipping the picture we currently have.

Finally, one of the shortcomings of our approach is that we cannot define
\texttt{POST} and \texttt{GET} endpoints independently
from each other and that we cannot explain \emph{middleware} as
a lens, in particular, some middlewares are allowed to perform state
updates including during the handling of \texttt{GET}
requests. Using the intuition from containers as ``question-answers''
mechanisms we are confident we can implement both those ideas.

    \bibliographystyle{alpha}
    \bibliography{./bibliography.bib}
\end{document}